\documentclass[sigplan,screen]{acmart}
\setcopyright{none}

\usepackage{booktabs}   
\usepackage{subcaption} 

\usepackage[utf8]{inputenc}
\usepackage{listings}
\usepackage{balance}
\usepackage{centernot}
\usepackage{xspace}
\usepackage[inference]{semantic}
\usepackage{algorithm}
\usepackage{algpseudocode}
\usepackage{caption}
\usepackage{makecell}
\usepackage{multirow}
\usepackage{stmaryrd}
\usepackage{amsmath}
\usepackage{graphicx}
\usepackage{setspace}
\usepackage{microtype}
\usepackage{mathtools}

\usepackage{wrapfig}
\usepackage{mathpartir}

\algrenewcommand\algorithmicindent{1em}

\graphicspath{{./figures/}}

\definecolor{programs}{gray}{0.1}
\definecolor{keywords}{HTML}{204a87}
\definecolor{comments}{HTML}{8f5902}
\definecolor{strings}{HTML}{4e9a06}

\makeatletter
\lst@CCPutMacro
    \lst@ProcessOther {"22}{\lst@ifupquote \textquotedbl
                                     \else \char34\relax \fi}
    \@empty\z@\@empty
\makeatother

\lstnewenvironment{rcodebox}
{\lstset{upquote=true,xleftmargin=2em,language=Ruby,breaklines=true,numbers=left,stepnumber=1,firstnumber=1,numberfirstline=true, frame=single, numberstyle={\footnotesize\it\color{programs}}}
}
{}

\def\spmid{\ \mid \ }

\newcommand\judgementHead[1]{\hfill\fbox{\ensuremath{#1}}}
\newcommand\name{\textsc{RbSyn}\xspace}
\newcommand\corelang{\ensuremath{\lambda_{syn}}\xspace}
\newcommand\type{\ensuremath{\tau}\xspace}
\newcommand\val{\ensuremath{v}\xspace}
\newcommand\expr{\ensuremath{e}\xspace}

\newcommand\branch{\ensuremath{b}\xspace}
\newcommand\meth{\ensuremath{m}\xspace}
\newcommand\class{\ensuremath{A}\xspace}
\newcommand\classobj[1]{\ensuremath{\lbrack #1 \rbrack}\xspace}
\newcommand\program{\ensuremath{P}\xspace}
\newcommand\precond{\ensuremath{S}\xspace}
\newcommand\postcond{\ensuremath{Q}\xspace}
\newcommand\classtable{\ensuremath{CT}\xspace}
\newcommand\vnil{\texttt{nil}\xspace}
\newcommand\vtrue{\texttt{true}\xspace}
\newcommand\vfalse{\texttt{false}\xspace}
\newcommand\var{\ensuremath{x}\xspace}
\newcommand\eseq[2]{\ensuremath{#1 ; #2}\xspace}

\newcommand\elet[3]{\ensuremath{\textbf{\texttt{let}}\ #1 = #2\ \textbf{\texttt{in}}\ #3}\xspace}
\newcommand\eif[3]{\ensuremath{\textbf{\texttt{if}}\ #1\ \textbf{\texttt{then}}\ #2\ \textbf{\texttt{else}}\ #3}\xspace}
\newcommand\emethcall[3]{\ensuremath{#1.#2(#3)}\xspace}

\newcommand\eprog[3]{\ensuremath{\textbf{\texttt{def}}\ #1(#2) = #3}}

\newcommand\ehole{\ensuremath\square\xspace}
\newcommand\effhole{\ensuremath\Diamond\xspace}
\newcommand\eassert[1]{\ensuremath{\textbf{\texttt{assert}}\ #1}\xspace}

\newcommand\methtype{\ensuremath{\sigma}\xspace}
\newcommand\mthtype[2]{\ensuremath{#1 \rightarrow #2}\xspace}

\newcommand\rulename[1]{\textsc{#1}\xspace}
\newcommand\region{\ensuremath{r}\xspace}
\newcommand\pure{\ensuremath{\bullet}\xspace}

\newcommand\code[1]{\texttt{#1}}
\newcommand\tenv{\ensuremath{\Gamma}\xspace}
\newcommand\eff{\ensuremath{\epsilon}\xspace}
\newcommand\err[2]{\ensuremath{\textbf{\texttt{err}}(#1, #2)}\xspace}
\newcommand\btime[2]{#1 \scriptsize{\ensuremath{\pm} #2}}
\newcommand\rwpair[2]{\ensuremath{\langle #1, #2 \rangle}}
\newcommand\hashole{\texttt{evaluable}\xspace}

\newcommand\size{\texttt{size}\xspace}
\newcommand{\spec}{\ensuremath{s}\xspace}
\newcommand{\specs}{\Psi}
\newcommand{\goal}{G}
\newcommand{\xret}{\ensuremath{x_r}}

\begin{document}

\title[]{\name: Type- and Effect-Guided Program Synthesis}         


\author{Sankha Narayan Guria}
\affiliation{
  \institution{University of Maryland}            
  \city{College Park}
  \state{Maryland}
  \postcode{20742}
  \country{USA}                    
}
\email{sankha@cs.umd.edu}          

\author{Jeffrey S. Foster}
\affiliation{
  \institution{Tufts University}            
  \city{Medford}
  \state{Massachusetts}
  \postcode{20742}
  \country{USA}                    
}
\email{jfoster@cs.tufts.edu}          

\author{David Van Horn}
\affiliation{
  \institution{University of Maryland}            
  \city{College Park}
  \state{Maryland}
  \postcode{20742}
  \country{USA}                    
}
\email{dvanhorn@cs.umd.edu}          

\begin{abstract}
  In recent years, researchers have explored component-based
  synthesis, which aims to automatically construct programs that
  operate by composing calls to existing APIs. However, prior work has
  not considered efficient synthesis of methods with side effects,
  e.g., web app methods that update a database. In this paper, we
  introduce \name{}, a novel type- and effect-guided synthesis tool
  for Ruby. An \name{} synthesis goal is specified as the type for the
  target method and a series of test cases it must pass. \name{} works
  by recursively generating well-typed candidate method bodies whose
  write effects match the read effects of the test case
  assertions. After finding a set of candidates that separately
  satisfy each test, \name{} synthesizes a solution that branches to
  execute the correct candidate code under the appropriate
  conditions. We formalize \name{} on a core, object-oriented language
  \corelang{} and describe how the key ideas of the model are
  scaled-up in our implementation for Ruby. We evaluated \name{} on 19
  benchmarks, 12 of which come from popular, open-source Ruby apps. We
  found that \name{} synthesizes correct solutions for all benchmarks,
  with 15 benchmarks synthesizing in under 9 seconds, while the
  slowest benchmark takes 83 seconds. Using observed reads to guide
  synthesize is effective: using type-guidance alone times out on 10
  of 12 app benchmarks. We also found that using less precise effect
  annotations leads to worse synthesis performance. In summary, we
  believe type- and effect-guided synthesis is an important step
  forward in synthesis of effectful methods from test cases.
\end{abstract}

\ccsdesc[500]{Software and its engineering~Automatic programming}

\keywords{program synthesis, type and effect systems, Ruby}  

\maketitle

\section{Introduction}
\label{sec:introduction}

A key task in modern software development is writing code that
composes calls to existing APIs, such as from a
library or framework. \emph{Component-based synthesis} aims to carry
out this task automatically, and researchers have shown how to perform
component-based synthesis using SMT solvers~\citep{jha2010oracle}; how
to synthesize branch conditions~ \citep{perelman-tds}; and how to
perform synthesis given a very large number of
components~\citep{feng-componentsyn}.

This prior work guides the synthesis process using types or special
properties of the synthesis domain, which is critical to achieving
good performance. However, prior work does not explicitly consider
\emph{side effects}, which are pervasive in many domains. For example,
consider synthesizing a method that updates a database. Without
reasoning about effects---in this case, that the method body needs to
change the database---synthesis of such a method reduces to
brute-force search, limiting its performance.

In this paper, we address this issue by introducing
\name{},
a new tool for synthesizing
Ruby methods. In \name{}, the user specifies the desired method by its type signature and a series of test cases it must pass. \name{} then searches for a solution by enumerating candidates and checking them against the tests. The key novelty of \name{} is that the search is both \emph{type- and effect-guided}. Specifically, the search begins with a \emph{typed hole} tagged with the method's return type. Each step either replaces a typed hole with an expression of that type, possibly introducing more typed holes; inserts an \emph{effect hole}, annotated with a write effect that may be needed to satisfy a test assertion; or replaces an effect hole with an expression with the given write effect, possibly inserting another effect hole. Once this process finds a set of method bodies that cumulatively pass all tests, \name{} uses a novel merging strategy to construct a complete solution: It creates a method whose body branches among the conditions, executing the corresponding (passing) code, thus yielding a single method that passes all tests. (\S~\ref{sec:overview} gives a complete example of \name{}'s synthesis process.)

We formalize \name{} for \corelang, a core object-oriented language. The synthesis algorithm is comprised of three parts. The first part, type-guided synthesis, is similar to prior work \citep{osera2015type,frankle2016example,polikarpova-synquid}, but is geared towards imperative, object-oriented programs. The second part is \emph{effect-guided synthesis}, which tries to fill an effect hole $\effhole:\eff $ with an expression with effect $\eff$. In \corelang{}, an effect accesses a \emph{region} $A.r$, where $A$ is a class and $r$ is an uninterpreted identifier. For example, \code{Post.author} might indicate reading instance field \code{author} of class \code{Post}. This notion of effects balances precision and tractability: effects are precise enough to guide synthesis effectively, yet coarse enough that reasoning about them is simple. The last part of the synthesis algorithm synthesizes branch conditions to create a \emph{merged} program that combines solutions for individual tests into an overall solution for the complete problem. (\S~\ref{sec:formalism} discusses our formalism.)


Our implementation of \name{} is built on top of RDL, a Ruby type system~\citep{rdl-github}. Our implementation extends RDL to include effect annotations, including a \code{self} region to give more precise effect information in the presence of inheritance. Our implementation also makes use of RDL's \emph{type-level computations}~\citep{kazerounian-comptype} to provide precise typing during synthesis. Finally, when searching for solutions, our implementation heuristically prioritizes further exploration of candidates that are small and have passed more assertions. (\S~\ref{sec:implementation} describes our implementation.)

We evaluated \name{} on a suite of 19 benchmarks, including seven benchmarks we wrote and 12 benchmarks extracted from three widely used, open-source Ruby apps: Discourse, Gitlab, and Disaspora. For the former, we wrote our own specifications. For the latter, we used unit tests that came with the benchmarks. We found that \name{} synthesizes correct solutions for all benchmarks and does so quickly, taking less than 9 seconds each for 15 of the benchmarks, and 83 seconds for the slowest benchmark. Moreover, type- and effect-guidance is critical. Without it, a majority of the benchmarks time out after five minutes. Finally, we examine the tradeoff of effect precision versus performance. We found that restricting effects to class names only causes 3 benchmarks to time out, and restricting effects to only purity/impurity causes 10 benchmarks to time out. (\S~\ref{sec:evaluation} discusses the evaluation in detail.)

We believe that \name{} is an important step forward in synthesis of effectful methods from test cases.

\section{Overview}
\label{sec:overview}

\begin{figure}
\begin{rcodebox}
# User schema {name: Str, username: Str}(*\label{line:user-schema}*)
# Post schema {author: Str, title: Str, slug: Str}(*\label{line:post-schema}*)

define :update_post, "(Str, Str, {author: ?Str, title: (*\label{line:type-sig-start}*)
    ?Str, slug: ?Str}) -> Post", [User, Post] do(*\label{line:type-sig}*)
  spec "author can only change titles" do(*\label{line:spec1}*)
    setup {
      seed_db  # add some users and their posts to db
      @post = Post.create(author: 'author', slug:
        'hello-world', title: 'Hello World')
      update_post('author', 'hello-world', author:(*\label{line:call-func}*)
        'dummy', title: 'Foo Bar', slug: 'foobar')
    }
    postcond { |updated|
      assert { updated.id == @post.id }(*\label{line:first-assert}*)
      assert { updated.author == "author" }(*\label{line:read-effect}*)
      assert { updated.title == "Foo Bar" }(*\label{line:third-assert}*)
      assert { updated.slug == 'hello-world' }
    }
  end
  spec "other users cannot change anything" do(*\label{line:spec2}*)
    setup { ... # same setup as above except next line
      update_post('dummy', ...) # other args same
    }
    postcond { |updated| ... # same other three asserts
      assert { updated.title == "Hello World" }
    }
end end
\end{rcodebox}
\caption{Specification for \code{update\_post} method}
\label{fig:overview-spec}
\end{figure}



In this section, we illustrate \name{} by using it to synthesize a
method from a hypothetical web blogging app. This app makes heavy use
of ActiveRecord, a popular database access library for Ruby on Rails. It
is the ActiveRecord methods whose side effects \name{} uses to guide
synthesis.

Figure~\ref{fig:overview-spec} shows the synthesis problem. This
particular app includes database tables for users and posts. In
ActiveRecord, rows of these tables are represented as instances of
classes \code{User} and \code{Post}, respectively. For reference, the
table schemas are shown in lines \ref{line:user-schema} and
\ref{line:post-schema}. Each user has a \code{name} and \code{username}.
Each post has the \code{author}'s username, the post's \code{title}, and
a \code{slug}, used to compute a permalink.

The goal of this particular synthesis problem, given by the call to
\code{define}, is to create a method \code{update\_post} that allows
users to change the information about a post.
Lines~\ref{line:type-sig-start} and~\ref{line:type-sig} specify the
method's type signature in the format of RDL~\citep{rdl-github}, a Ruby
type system that \name{} uses for types and type checking. Here, the first two arguments are strings, and the last
is a \emph{finite hash type} that describes an instance of
\code{Hash} with optional (indicated by \code{?}) keys \code{author},
\code{title}, and \code{slug} (all \emph{symbols}, which are just
interned strings) that map to strings. The method itself returns a
\code{Post}.

In addition to the type signature, the synthesis problem also includes a
list of constants that can be used in the target method.
In this case, those constants are the classes \code{User} and \code{Post},
as given by the last argument to \code{define} on
line~\ref{line:type-sig}. These classes can then be used to invoke
singleton (class) methods in the synthesized method. For simplicity, we
assume that \name{} can use only these constants for this example. In
practice, \name can synthesize predefined numeric or string constants
like 0, 1 or the empty string.

Finally, the synthesis problem includes a number of \emph{specs}, which are just test cases. Each spec has a title, for human convenience; a \code{setup} block to establish
any necessary preconditions and call the synthesized
method; and a \code{postcond} block with assertions that must hold after
the synthesized method runs. As we will see below, separating the pre-
and postconditions allows \name{} to more easily use effects to guide
synthesis. In this example, both specs add a few users and a post
created by each of them to the
database (call to \code{seed\_db}, details not shown) and then create a
post titled ``Hello World'' by the user \code{author}. The first spec
asserts that \code{update\_post} allows \code{author} to update a post's
title. The second spec asserts that a \code{dummy} user 
cannot update the post. The check for \code{id}
ensures that only existing posts are updated (any new posts will have a
new unique id).

The final, synthesized solution is shown on the right of Figure~\ref{fig:overview-diagram-merged-prog}. Notice the synthesized code calls several
ActiveRecord methods (\code{exists?}, \code{where}, and \code{first}) as
well as the hash access method \code{[]}. Applying solver-aided
synthesis to this problem would require developing accurate models of these methods, which is a difficult
challenge~\cite{nijjar2011bounded}. To address this limitation, \name{}
instead enumerates candidates, which can then be run to check them
against the specs. As the search space is vast, \name{} uses
\code{update\_post}'s type signature and the effects from the specs'
\code{postcond}s the guide the search. Finally, \name{} uses a novel
merging algorithm to synthesize the necessary branch condition to yield
a solution that satisfies both specs.



\subsection{Synthesizing Spec Solutions}
\label{subsec:overview-type-directed}

\begin{figure*}
\centering
\begin{subfigure}[b]{0.66\textwidth}
  \centering
  \includegraphics[scale=0.42]{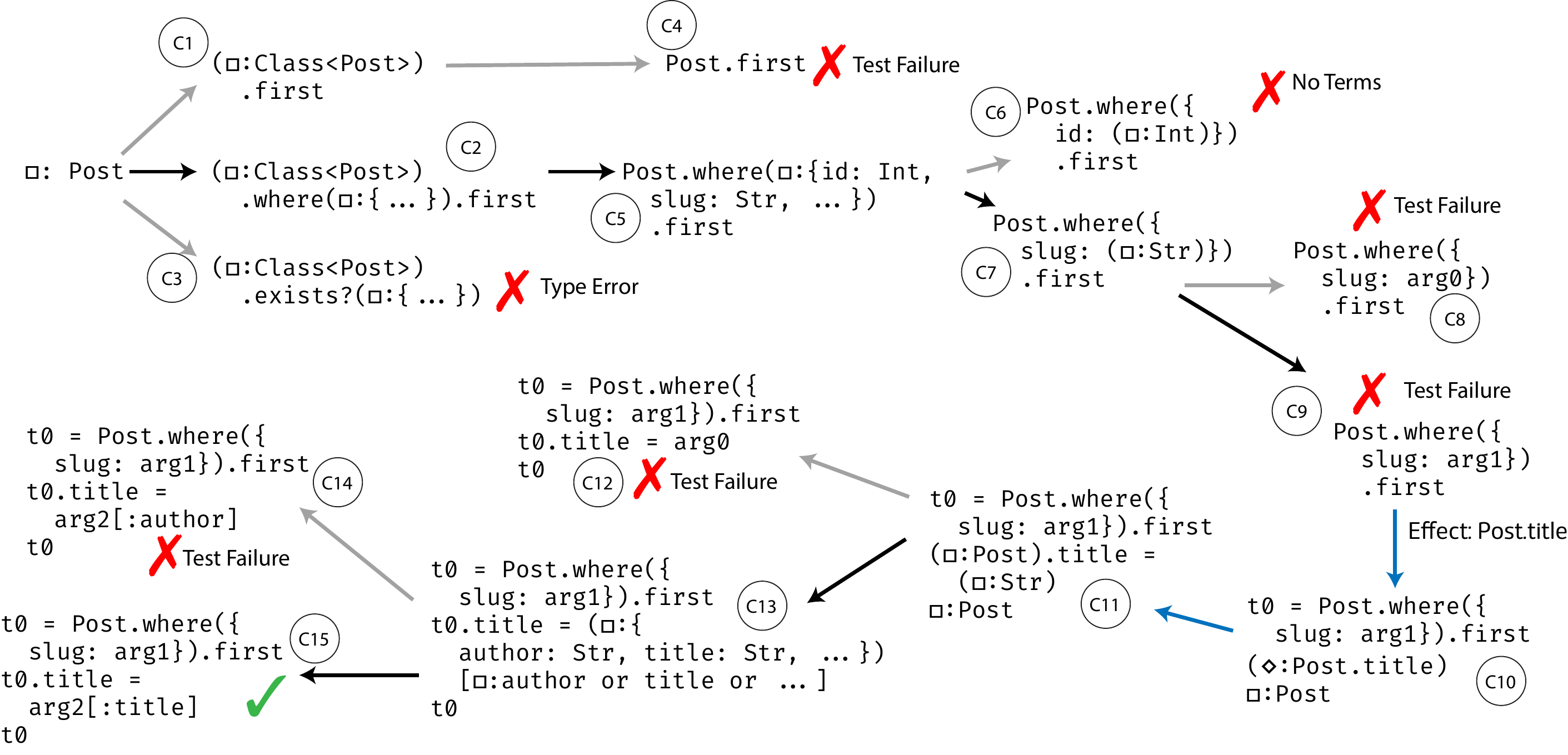}
\end{subfigure}
\hfill
\begin{subfigure}[b]{0.32\textwidth}
  \begin{rcodebox}
def update_post(arg0, arg1, arg2)
  if Post.exists?(author: arg0,(*\label{line:prog4-start}*)
      slug: arg1)
    t0 = Post.where(slug:arg1).first(*\label{line:prog1-start}*)
    t0.title=arg2[:title]
    t0(*\label{line:prog1-end}*)
  else
    Post.where(slug: arg1).first(*\label{line:prog2}*)
  end(*\label{line:prog4-end}*)
end
  \end{rcodebox}
\end{subfigure}
\caption{\emph{Left:} Steps in the synthesis of solution to the first
specification. Note C2 takes two steps to synthesize but is shown as a single composite step.
Some choices available to the synthesis algorithm have been omitted for
simplicity. \emph{Right:} Synthesized \code{update\_post} method.}
\label{fig:overview-diagram-merged-prog}
\end{figure*}

The left portion of Figure~\ref{fig:overview-diagram-merged-prog} shows the search process
\name{} uses to solve this synthesis problem. To begin, \name{} observes
that the return type of \code{update\_post} is \code{Post}. Thus, the
search begins (upper left) by creating a candidate method body
\code{\ehole:Post}, which is a \emph{typed hole} that must be filled by an
expression of type \code{Post}. \name{} then iteratively expands holes
in candidates, running the specs whenever it produces
fully concretized candidates with no holes.

In general, \name{} can fill a typed hole with a local variable, a constant,
or a method call. As there are no local variables
(which so far are just parameters) or constants of the
appropriate type, 
\name{} chooses a method call. To do so, it searches
through the available method type annotations to find those that could
return \code{Post}. In this case, \name{} takes advantage of RDL's type
annotations for ActiveRecord \citep{kazerounian-comptype} to synthesize
candidates \textsf{C1} and \textsf{C2}, among others (not shown).
It is straightforward for the user to add type annotations for any other
library methods that might be needed by the synthesized method. For
illustration purposes, we also show a candidate \textsf{C3} that returns
the wrong type. Such candidates are discarded by \name{}, vastly
reducing the search space. Note that C2 contains two method calls, and thus would take two steps to produce, but we show it here as a single step for conciseness.

Next, \name{} tries to fill holes in candidate expressions,
starting with smaller candidates. In this case, it first considers \textsf{C1}, which has a hole of type
\code{Class<Post>}, which is the singleton type for the constant
\code{Post}. Thus, there is
only one choice for the hole, yielding candidate \textsf{C4}. Since
\textsf{C4} has no holes, \name{} runs it against the specs. More
specifically, it runs it against the first spec---as we will discuss
shortly, \name{} synthesizes solutions for each spec independently, and
then combines them. In this case, \textsf{C4} fails the spec (because
the first post in the database is not the one to be updated, due to the
initial database seeding) and hence is rejected.

Continuing with \textsf{C5}, \name{} fills in the (finite
hash-typed) hole, yielding choices that include \textsf{C6} and
\textsf{C7}. \name rejects \textsf{C6} since there is no way to
construct an expression of type \code{Int}.
However, for \textsf{C7}, there are two local variables of type
\code{Str} from the method arguments. Substituting these yields
\textsf{C8} and \textsf{C9}. \textsf{C8} uses \code{arg0}, the username,
to query the \code{Post} table's slug, so it fails. \textsf{C9} queries
the \code{Post} table with the correct slug value \code{arg1}. This
passes the first two assertions (line~\ref{line:first-assert} onwards)
but fails the third, which expects the post title to be updated
from ``Hello World'' to ``Foo Bar.''

\name{} extends RDL's type annotations to include read and write effects. When the expression inside an \code{assert} evaluates to \code{false}, \name{} infers the \code{assert}'s read and write effects based on those of the methods it calls. For example, we can give the
\code{Post\#title}\footnote{\code{A\#m} indicates instance method
\code{m} of class \code{A}.} method, used by the third assertion, the
following signature:
\begin{lstlisting}
type Post, :title, '() -> Str', read: ['Post.title']
\end{lstlisting}
Thus, \name{} sees that the failing assertion reads
\code{Post.title}, an abstract effect label. To make the assertion
succeed, \name{} inserts an \emph{effect hole} $\effhole:\code{Post.title}$ in the candidate program (\textsf{C10}). It also saves the value of the previous candidate expression in a temporary variable, and inserts a hole with the candidate's type at the end. \name{} then continues the search, trying to fill the effect hole with a call to a method whose \emph{write} effect matches the hole---such a call could potentially satisfy the failed assertion. Here, \name{} replaces the effect hole (\textsf{C11}) with a call to \code{Post\#title=}, which is such a method.
(We should note that all previous candidates that failed a spec due to a side effect will also have effect holes added in a similar fashion. We omit these candidates from the discussion as they do not lead to a solution.)


\name{} continues by using type-guided synthesis for the typed holes of
\textsf{C11}---yielding \textsf{C12}, rejected due to assertion
failures---and then \textsf{C13}. After several steps (not shown),
\name{} arrives at \textsf{C14}, which fails the spec, and \textsf{C15},
which fully satisfies the first spec. Indeed, we see this exact
expression in lines~\ref{line:prog1-start}--\ref{line:prog1-end} of the
solution in Figure~\ref{fig:overview-diagram-merged-prog}.


\subsection{Merging Solutions}

\name{} next uses the same technique to synthesize an expression that
satisfies the second spec, yielding the expression shown on
line~\ref{line:prog2}. Now \name{} needs
to merge these individual solutions into a single solution that passes
all specs. At a high-level, it does so by constructing a program
\code{if $\branch_1$ then $\expr_1$ else if $\branch_2$ then $\expr_2$ end}, where the \code{$\expr_i$} are the
solutions for the specs and the \code{$\branch_i$} are \emph{branch conditions}
capturing the conditions under which those expressions pass the specs.

To create the \code{$\branch_i$}, \name{} uses the same technique again, this
time synthesizing a boolean-valued expression that evaluates to
\code{true} under the \code{setup} of spec $i$. In this case,
this process results in the same branch condition \code{\vtrue} for both
specs. However, since this trivially holds for both specs,
this branch condition does not work---we need to find a branch condition
that distinguishes the two cases.

Next \name{} tries to synthesize a branch condition \code{$\branch'_1$}
that evaluates to \code{true} for the \code{setup} of the first spec and
\code{false} for the \code{setup} of the second. This yields the more precise branch condition
\code{$\branch'_1$ = Post.exists?(author: arg0, slug: arg1)}. This is a sufficient condition, as
the
\code{update\_\-post} method is supposed to update a post only if a post
with slug \code{arg1} is authored by \code{arg0}.
It solves an analogous
synthesis problem for the second spec, yielding \code{$\branch'_2$ =
!Post.\-exists?(author: arg0, slug: arg1)}. As these are the negation of each
other, \name{} then merges these two together as \code{if-then-else}
(rather than an \code{if-then-else if-\-then-else}), yielding the final
synthesized program in Figure~\ref{fig:overview-diagram-merged-prog}.



\section{Formalism}
\label{sec:formalism}

\begin{figure}
  \centering
  
  $$
  \begin{array}{lccl}
  \emph{Values} \quad
  & \val
  & ::= & \vnil \spmid \vtrue \spmid \vfalse \spmid \classobj{\class} \\
  \emph{Expressions} \quad
  & \expr
  & ::= & \val \spmid \var \spmid \eseq{\expr}{\expr} \spmid \emethcall{\expr}{\meth}{\expr} \\
  &     & \spmid & \eif{\branch}{\expr}{\expr} \\
  &     & \spmid & \elet{\var}{\expr}{\expr} \spmid \ehole: \type \spmid \effhole: \eff \\
  \emph{Conditionals} \quad
  & \branch
  & ::= & \expr \spmid $!\branch$ \spmid \branch\lor\branch\\
  \emph{Types} \quad
  & \type
  & ::= & \class \spmid \type \cup \type \\
  \emph{Programs} \quad
  & \program
  & ::= & \eprog{\meth}{\var}{\expr} \\
  \\

  \emph{Specs} \quad
  & \spec
  & ::= & \langle \precond, \postcond \rangle \\
  \emph{Setup} \quad
  & \precond
  & ::= & \eseq{\expr}{\;\xret=\program(\expr)} \\
  \emph{Postconditions} \quad
  & \postcond
  & ::= & \eassert{\expr} \spmid \eseq{\postcond}
  {\postcond} \\
  \emph{Spec Set} \quad
  & \specs
  & := & \{s_i\} \\
  \emph{Synthesis Goal} \quad
  & \goal
  & ::= & \langle \type \rightarrow \type, \specs \rangle \\
  \\
  
  \emph{Class Table} \quad
  & \classtable
  & ::= & \emptyset \spmid \class.\meth: \methtype, \classtable\\
  \emph{Method Types} \quad
  & \methtype
  & ::= & \type \xrightarrow[]{\rwpair{\eff_r}{\eff_w}} \type \\
  \emph{Type Env.} \quad
  & \tenv
  & ::= & \emptyset \spmid \var: \type, \tenv \\
  \emph{Dynamic Env.} \quad
  & E
  & ::= & \var \rightarrow \val \\
  \emph{Constants} \quad
  & \Sigma
  & ::= & \emptyset \spmid \val: \type, \Sigma \\

  \emph{Effect} \quad
  & \eff
  & ::= & \pure \spmid * \spmid \class.* \spmid \class.\region \spmid \eff \cup \eff\\
  \end{array}
  $$

  $\region \in \textrm{effect regions}
  \quad
  \pure \subseteq \eff
  \quad
  \eff \subseteq *
  $

  $\class_1.* \subseteq \class_2.* \textnormal{ and } \class_1.\region \subseteq \class_2.\region \textnormal{ and } \class_1.\region \subseteq \class_2.* \textnormal{ if}\ \class_1 \leq \class_2$

  $
  \eff^1 \subseteq \eff^1 \cup \eff^2
  \quad
  \eff^2 \subseteq \eff^1 \cup \eff^2
  $

  $
  \rwpair{\eff^1_r}{\eff^1_w} \cup \rwpair{\eff^2_r}{\eff^2_w} =
  \rwpair{\eff^1_r \cup \eff^2_r}{\eff^1_w \cup \eff^2_w}
  $

	\bigskip{}

  $\var \in \textrm{variables}$,
  $\meth \in \textrm{methods}$,
  $\class \in \textrm{classes}$,\\

  $\emph{Nil} \leq \type \quad \type \leq \emph{Obj} \quad
  \type_1 \leq \type_1 \cup \type_2 \quad \type_2 \leq \type_1
  \cup \type_2$\\

  \caption{Syntax and Relations of \corelang.}
  \label{fig:lang}
\end{figure}

In this section, we formalize \name on \corelang, a core object-oriented calculus shown in Figure~\ref{fig:lang}. Values \val include
\vnil, \vtrue, \vfalse, and objects \classobj{\class} of class \class.
Note that we omit fields to keep the presentation simpler. Expressions
\expr include values, variables \var, sequences \eseq{\expr}{\expr},
method calls \emethcall{\expr}{\meth}{\expr}, conditionals
\eif{\branch}{\expr}{\expr}, and variable bindings
\elet{\var}{\expr}{\expr}. A conditional guard \branch can be an
expression \expr, a negation $!\branch$, or a disjunction
$\branch\lor\branch$. The grammar for guards is limited to match what
\name{} can actually synthesize.

Expressions also include \emph{typed
holes} $\ehole: \type$ and \emph{effect holes}
$\effhole: \eff$, which are placeholders that are eventually
filled with an expression of the given type, or expression with the
given write effect, respectively. 
We note our synthesis algorithm only inserts effect holes at positions that can have any type.
Types are either classes or unions of types, and we assume classes form a  lattice with \emph{Nil} (the class of \vnil) as the bottom element and \emph{Obj} as the top element. We write $A \leq B$ when class $A$ is a subclass of $B$ according to the lattice.
We defer the definition of effects for the moment. Finally, a synthesized program \program is a single method
definition \eprog{\meth}{\var}{\expr}. We restrict the method to one
argument for convenience.

A spec \spec{} in \corelang{} is a pair of setup code \precond and a
postcondition \postcond. A setup $\eseq{\expr_1}{\xret=\program(\expr_2)}$
includes some initialization $\expr_1$ followed by a special form
indicating calling the synthesized method in
$P$ with argument $\expr_2$ and binding the result to $\xret$. The
postcondition is a sequence of assertions that can test $\xret$ and
inspect the global state using library methods. We write $\specs$ for a set of specs, and a \emph{synthesis goal} $\goal$ is a pair $\langle \type_1\rightarrow\type_2, \specs\rangle$, where $\type_1$ and $\type_2$ are the method's domain and range types, respectively, and $\specs$ are the specs the synthesized method should satisfy.



The next part of Figure~\ref{fig:lang} defines additional notation used in the formalism. Synthesized methods can use classes and methods from a \emph{class table} \classtable, which maps class and method names to the methods' types. For example, the class table has type information for other methods of a target app and library methods such as those from ActiveRecord. A method type $\methtype$ has the form $\type \xrightarrow[]{\rwpair{\eff_r}{\eff_w}} \type'$, where $\type$ and $\type'$ are the domain and range types, respectively, and \rwpair{\eff_r}{\eff_w} specifies the method's read effect $\eff_r$ and write effect $\eff_w$ (discussed shortly). During type-guided synthesis, \name maintains a type environment \tenv mapping variables to their types. When executing a synthesized program, the operational semantics (omitted) uses a dynamic environment $E$ mapping variables to their values. During synthesis, $\Sigma$ is a list of user-supplied constants that can fill holes.




%
%
%
%
%

\paragraph*{Effects.}

The last part of Figure~\ref{fig:lang} defines effects $\eff$. In \name{}, effects are hierarchical names that abstractly label the program state. The empty effect \pure denotes no side effect, used for pure computations. The effect $*$ is the top effect, indicating a computation that might touch any state in the program. Lastly, effect $\class.*$ denotes code that touches any state within class $\class$, and $\class.\region$ denotes code that touches the region labeled \region in $\class$, where region names are completely abstract. Effects can also be unioned together.

We define subsumption $\eff_1 \subseteq \eff_2$ on effects to hold when $\eff_2$ includes $\eff_1$. Effects \pure and $*$ are the bottom and top, respectively, of the $\subseteq$ relation, and if $A_1 \leq A_2$ then $A_1.\region \subseteq A_2.\region$ and $A_1.\region \subseteq A_2.*$ and $A_1.* \subseteq A_2.*$. We also have standard rules for subsumption with effect unions.

In \name{}, all effects arise from calling methods from the class table $\classtable$, which have effect annotations of the form \rwpair{\eff_r}{\eff_w}, where $\eff_r$ and $\eff_w$ are the method's read and write effects, respectively. We extend subsumption to such paired effects in the natural way. During synthesis, if \name{} observes the failure of an assertion with some read effect $\eff_r$, it tries to fix the failure by inserting a call to some method with write effect $\eff_w$ such that $\eff_r \subseteq \eff_w$, i.e., it tries writing to the state that is read. For example, in Section~\ref{sec:overview}, this technique generated a call to \code{Post\#title}. 

Our effect language is inspired by the region path lists approach of
\citet{bocchino-rpl}, but is much simpler. We opted for
coarse-grained, abstract effects to make it easier to write
annotations for library methods. Although class names are included in
the effect language, such names are for human convenience
only---nothing precludes a method in class $A$ being annotated with an effect to $B.\region$ for some other class $B$. We found that this approach works well for our problem setting of synthesizing code for Ruby apps, where trying to precisely model heap and database state would be difficult. However, we believe the core of this approach---pairing effects (in our case, reads and writes) and then creating candidates using the opposing element of such a pair---can be generalized to more complex effect systems.

\paragraph*{Synthesis Problem.}

We can now formally specify the synthesis problem. Given a synthesis
goal $\langle \type_1 \rightarrow \type_2, \{ \langle \precond_i,
\postcond_i \rangle\} \rangle$, \name{} searches for a program $P$ such
that, for all $i$, assuming that $\precond_i$ calls $P$ with an argument
of type $\type_1$, evaluating to $\var_r$ of type $\type_2$, it is the case that $P \vdash
\eseq{\precond_i}{\postcond_i} \Downarrow \val$. In other words, evaluating the setup followed by the
postcondition yields some value rather than aborting
with a failed assertion. We omit the evaluation rules
as they are standard.

\subsection{Type-Guided Synthesis}
\label{subsec:type-directed}

\begin{figure}
\centering
\judgementHead{\Sigma, \tenv \vdash_{\classtable}
\expr \rightsquigarrow \expr: \type}
\begin{mathpar}
\inference{\tenv(\var) = \type}{\Sigma, \tenv \vdash_
{\classtable} \var \rightsquigarrow \var: \type}[
\rulename{T-Var}]

\inference{\Sigma, \tenv \vdash_{\classtable} \expr_1
\rightsquigarrow \expr'_1: \type_1\\
\Sigma, \tenv[\var \mapsto \type_1] \vdash_{\classtable} \expr_2
\rightsquigarrow \expr'_2: \type_2}
{\Sigma, \tenv \vdash_{\classtable}
\elet{\var}{\expr_1}{\expr_2} \rightsquigarrow
\elet{\var}{\expr'_1}{\expr'_2}: \type_2}[\rulename{T-Let}]

\inference{}{\Sigma, \tenv \vdash_{\classtable}
\ehole: \type \rightsquigarrow (\ehole: \type) : \type}[\rulename{T-Hole}]

\inference{\val: \type_1 \in \Sigma &
\type_1 \leq \type_2}
{\Sigma, \tenv \vdash_{\classtable} \ehole: \type_2
\rightsquigarrow \val: \type_1}[\rulename{S-Const}]

\inference{\tenv(\var) = \type_1 &
\type_1 \leq \type_2}
{\Sigma, \tenv \vdash_{\classtable} \ehole: \type_2
\rightsquigarrow \var: \type_1}[\rulename{S-Var}]

\inference{\meth: \mthtype{\type_1}{\type_2} \in \classtable(\class) &
\type_2 \leq \type_3}
{\Sigma, \tenv \vdash_{\classtable} \ehole:\type_3
\rightsquigarrow \emethcall{(\ehole: \class)}{\meth}{\ehole: \type_1}: \type_2}[\rulename{S-App}]

\end{mathpar}
\caption{Type-guided synthesis rules (selected).}
\label{fig:type-synthesis-rules}
\end{figure}

The first component of \name{} is type-guided synthesis, which creates candidate expressions of a given type by trying to fill a hole $\ehole:\type_2$ where $\type_2$ is the method return type. Figure~\ref{fig:type-synthesis-rules} shows a subset of the type-guided synthesis rules; the full set can be found in Appendix~\ref{subsec:appendix-type-syn}. These rules have the form $\Sigma, \tenv \vdash_{\classtable} \expr_1 \rightsquigarrow \expr_2: \type$, meaning with constants $\Sigma$, in type environment $\tenv$, under class table $\classtable$, the holes in  $\expr_1$ can be rewritten to yield $\expr_2$, which has type $\type$.

The rules in Figure~\ref{fig:type-synthesis-rules} have two forms. The \rulename{T-} rules apply to expressions whose  outermost form is not rewritten. Thus these rules perform standard type checking. For example, \rulename{T-Var} type checks a variable \var by checking its type against the type environment \tenv, leaving the term unchanged. \rulename{T-Let} typechecks and recursively rewrites (or not) the subexpressions and then  rewrites those new expressions into a let-binding, ensuring the resulting term is type-correct. Finally, \rulename{T-Hole} applies to a typed hole that is not being rewritten, in which case it remains the same and has the given type.

The \rulename{S-} rules rewrite typed holes. \rulename{S-Const} replaces a hole by a constant of the correct type from $\Sigma$. \rulename{S-Var} is similar, replacing a hole by a variable from $\tenv$. Finally, \rulename{S-App} replaces a hole with a call to a method with the right return type, inserting typed holes for the method receiver and argument.

\paragraph*{Type Narrowing.}
Notice that in these three S-rules, the term replacing the hole may
actually have a subtype of the original hole's type. Thus, type-guided
synthesis could \emph{narrow} types in a synthesized
program, potentially also narrowing the search space. For example, consider an
expression $\emethcall{(\ehole_1:\code{Str})}{\code{append}}{\ehole_2:\code{Str}}$ that joins two strings, and assume the set of constants $\Sigma$
includes \vnil. Notice that \vnil is a valid substitution for $\ehole_1$, which will then cause the type of the receiver to narrow to \emph{Nil}. But then the typing derivation fails because the \emph{Nil} type
has no \code{append} method, stopping further exploration along this path. In contrast, if we had typed the replacement term at \code{Str}, then \name{} would have fruitlessly continued the search, trying various replacements for $\ehole_2$ only to reject them due to a runtime failure for invoking a method on $\vnil$.

\subsection{Effect-Guided Synthesis}
\label{subsec:effect-guided}


\begin{figure}
\centering
\judgementHead{\Sigma, \tenv, \eff_r \vdash_{\classtable}
\expr \twoheadrightarrow \expr}
\begin{mathpar}
\inference{\Sigma, \tenv \vdash_{\classtable} \expr
\rightsquigarrow \expr: \type}
{\Sigma, \tenv, \eff_r \vdash_{\classtable}
\expr \twoheadrightarrow
\elet{\var}{\expr}{(\eseq{\effhole: \eff_r}{\ehole:
\type})}}[\rulename{S-Eff}]
\end{mathpar}

\bigskip{}

\judgementHead{\Sigma, \tenv \vdash_{\classtable}
\expr \rightsquigarrow \expr: \type}
\begin{mathpar}
\inference{ }
{\Sigma, \tenv \vdash_{\classtable} \effhole: \eff \rightsquigarrow (\effhole: \eff): \emph{Obj}}
	[\rulename{T-EffObj}]

\inference{\eff_r \subseteq \eff'_w &
\meth: \type_1 \xrightarrow[]{\rwpair{\eff'_r}{\eff'_w}} \type_2 \in \classtable(\class)}
{\Sigma, \tenv \vdash_{\classtable}
\effhole: \eff_r \rightsquigarrow
\eseq{\ehole: \eff'_r}{\emethcall{(\ehole: \class)}{\meth}{\ehole: \type_1} : \type_2}}[\rulename{S-EffApp}]

\inference{ }
{\Sigma, \tenv \vdash_{\classtable} \effhole: \eff \rightsquigarrow \vnil: \emph{Nil}}
	[\rulename{S-EffNil}]
\end{mathpar}
\caption{Effect guided synthesis rule}
\label{fig:effect-synthesis-rules}
\end{figure}

The second component of \name is effect-guided synthesis, used when
type-guided synthesis creates a candidate that does not satisfy the
postcondition of the tests. If this happens, \name computes the effect
\rwpair{\eff_r}{\eff_w} of the failed assertion in the postcondition.
(We defer the formal rules for computing this effect to
Appendix~\ref{subsec:evaluation-rules}, as they simply
union the effects of method calls in the assertion.) Then, we
hypothesize that the assertion may have failed because the region
denoted by $\eff_r$ is in the wrong state.

To potentially fix the state, \name applies a new rule
\rulename{S-Eff}, shown in
Figure~\ref{fig:effect-synthesis-rules}. The hypothesis computes the
type $\type$ of $e$, the candidate expression that failed the
postcondition. In the conclusion, $e$ is rewritten to
$\elet{\var}{\expr}{(\eseq{\effhole: \eff_r}{\ehole: \type})}$, i.e.,
$e$ is computed, bound to $x$, and two holes are sequenced. The first
must be filled with an expression of the desired effect $\eff_r$. The second must have $e$'s type $\tau$, to preserve type-correctness. For example, it could be filled by $x$, as happened in Figure~\ref{fig:overview-diagram-merged-prog} when \code{t0} is returned.

The rules for working with effect holes are shown in the bottom of
Figure~\ref{fig:effect-synthesis-rules}, which extends
Figure~\ref{fig:type-synthesis-rules}. \rulename{T-EffObj} gives an
effect hole, that is not rewritten, type \emph{Obj}. Since this is the
top of the type hierarchy, this ensures an effect hole can safely be
replaced by a term with any type. In other words, effect holes are
filled for their effects, not their types. \rulename{S-EffApp} does the
heavy lifting, filling an effect hole with a call to a method $m$ with a
write effect $\eff'_w$ that subsumes the desired effect $\eff_r$. Of
course, this call may itself read state $\eff'_r$, so the rule precedes
the method call with a hole with that effect, in case said state needs
to change. Finally, \rulename{S-EffNil} replaces an effect hole with
$\vnil$, which removes it from the program. This is used in case some
extra effect holes are added that are not actually needed.

\subsection{Merging Solutions}
\label{subsec:merge-progs}

The last component of \name{} combines expressions that pass individual
specs into a final program that passes all specs. More specifically,
given a synthesis goal $\langle \type_1 \rightarrow \type_2, \{ \spec_i
\} \rangle$, \name{} first uses type- and effect-guided synthesis to
create expressions $e_i$ such that $e_i$ is the solution for spec
$\spec_i$. Then, \name{} combines the $e_i$ into a branching program
roughly of the form $\textbf{\texttt{if}}\ b_1\ \textbf{\texttt{then}}\
e_1\ \textbf{\texttt{else if}}\ b_2\ \textbf{\texttt{then}}\ e_2 \ldots$
for some $b_i$.



For each $i$, \name{} uses the type-guided synthesis rules in
\S~\ref{subsec:type-directed} to synthesize a $b_i$ such
that under the setup $\precond_i$ of spec $\spec_i$, conditional
$\branch_i$ evaluates to \vtrue, i.e., $\eprog{\meth}{\var}{\branch_i} \vdash
\precond_i; \eassert{\var_r} \Downarrow \val$. Note effect-guided synthesis is not used here as the asserted expression $\var_r$ is pure.

Notice that while each initial $b_i$ evaluates to $\vtrue$ under the precondition, there is no guarantee it is a sufficient condition for $\spec_i$ to satisfy the postcondition---especially because \name{} aims to synthesize small expressions, as discussed further in \S~\ref{sec:implementation}. Moreover, there may be multiple $e_i$ that are actually the same expression, and therefore could be combined to yield a smaller solution.

\begin{figure}
\begin{align}
\begin{split}
\langle \expr_1, \branch_1, \specs_1 \rangle \oplus \langle \expr_2, \branch_2, \specs_2 \rangle = \langle \expr_1, \branch_1, \specs_1 \cup \specs_2 \rangle\\
\textnormal{if}\ \expr_1 \equiv \expr_2\ \textnormal{and}\ \branch_1 \implies \branch_2 \label{eq:merge-rule2}
\end{split}
\\
\begin{split}
\langle \expr_1, \branch_1, \specs_1 \rangle \oplus \langle \expr_2, \branch_2, \specs_2 \rangle = \langle \expr_1, \branch_1 \lor \branch_2, \specs_1 \cup \specs_2 \rangle\\
\textnormal{if}\ \expr_1 \equiv \expr_2\ \textnormal{and}\ \branch_1 \centernot\implies \branch_2 \label{eq:merge-rule3}
\end{split}
\\
\begin{split}
\langle \expr_1, \branch_1, \specs_1 \rangle \oplus \langle \expr_2, \branch_2, \specs_2 \rangle = \langle \expr_1, \branch^{syn}_1, \specs_1 \rangle \oplus \langle \expr_2, \branch^{syn}_2, \specs_2 \rangle\\
\textnormal{if}\ \expr_1 \not\equiv \expr_2\ \textnormal{and}\ \branch_1 \implies \branch_2\\
\textnormal{where}\ \forall \langle \precond_i, \postcond_i \rangle
\in \specs_1. \eprog{\meth}{\var}{\branch^{syn}_1} \vdash \precond_i; \eassert{\var_r} \Downarrow \val\\
\land \quad \forall \langle \precond_j, \postcond_j \rangle \in \specs_2.
\eprog{\meth}{\var}{\branch^{syn}_1} \vdash \precond_j; \eassert{!\var_r} \Downarrow \val\\
\textnormal{and}\ \forall \langle \precond_i, \postcond_i \rangle \in
\specs_1. \eprog{\meth}{\var}{\branch^{syn}_2} \vdash \precond_i; \eassert{!\var_r} \Downarrow \val\\
\land \quad \forall \langle \precond_j, \postcond_j \rangle \in \specs_2.
\eprog{\meth}{\var}{\branch^{syn}_2} \vdash \precond_j; \eassert{\var_r} \Downarrow \val \label{eq:merge-rule4}
\end{split}
\end{align}
\caption{Rewriting rules.}
\label{fig:term-rewriting}
\end{figure}

Thus, \name{} next performs a \emph{merging} step to create the final
solution. This process operates on tuples of the form $\langle \expr,
\branch, \specs \rangle$, which is a hypothesis that the program
fragment $\textbf{\texttt{if}}\ \branch\ \textbf{\texttt{then}}\ \expr$
satisfies the specs $\specs$. \name{} repeatedly merges such tuples
using an operation $\langle \expr_1, \branch_1,
\specs_1 \rangle \oplus \langle \expr_2, \branch_2, \specs_2 \rangle$
to represent that
$\eif{\branch_1}{\expr_1}{\textbf{\texttt{if}}\ \branch_2\
\textbf{\texttt{then}}\ \expr_2}$ satisfies the specs $\specs_1 \cup
\specs_2$. We define $\textsc{Specs}(\langle \expr_1, \branch_1, \specs_1 \rangle
\oplus ...) = \bigcup \specs_i$, i.e., the specs from merged tuples, and
$\textsc{Prog}(\langle \expr_1, \branch_1, \specs_1 \rangle \oplus ...)$
$=\eprog{\meth}{\var}{\eif{\branch_1}{\expr_1}{...}}$, a definition with the expression represented by the merged tuples.

Figure~\ref{fig:term-rewriting} defines rewriting rules that are
applied to create the final solution.
Rule~\ref{eq:merge-rule2} simplifies the case where $\expr_1$ and
$\expr_2$ are the same and $\branch_1$ implies $\branch_2$, yielding a
single expression and branch that satisfy $\specs_1 \cup \specs_2$. Note
we omit the symmetric case for all rules due to space limitations.
Rule~\ref{eq:merge-rule3} applies when $\branch_1$ does not imply
$\branch_2$ but $\expr_1$ and $\expr_2$ are the same. In this case,
$\expr_1$ satisfies the union of the specs under the disjunction of the
branch conditions. (Note this rule could also applied if $\branch_1
\Rightarrow \branch_2$, but the resulting solution would be longer than
Rule~\ref{eq:merge-rule2} generates.) Finally, Rule~\ref{eq:merge-rule4}
applies when $\expr_1$ and $\expr_2$ differ but $\branch_1$ implies
$\branch_2$. In such a scenario, $\branch_2$ holds for both $\expr_1$ and
$\expr_2$ and thus it must be that $\branch_1$ and $\branch_2$ are
insufficient to branch among $\expr_1$ and $\expr_2$. Thus, \name{}
synthesizes a stronger conditional $\branch^{syn}_1$ that
hold for all specs in $\specs_1$ and does not hold for the specs in
$\specs_2$, and the reverse for $\branch^{syn}_2$. For example, recall the
application of this rule in the example of \S~\ref{sec:overview}, to
synthesize a more precise branch condition because the initial condition
\vtrue was the same for both branches.

\name{} also includes a number of other merging rules, deferred to
Appendix~\ref{subsec:branch-pruning}, for further simplifying
expressions. 
For example, 
 \eif{\branch_1}{\expr_1}{\eif{!\branch_1}{\expr_2}{\vnil}} can be rewritten as \eif{\branch_1}{\expr_1}{\expr_2}, which was used to generate the solution in Figure~\ref{fig:overview-diagram-merged-prog}.

\paragraph*{Checking Implication.}

Checking the implications in Figure~\ref{fig:term-rewriting} is
challenging since branch conditions may include method calls whose
semantics is hard to reason about. To solve this problem, \name{} checks
implications using a heuristic approach that is effective in practice. Each
unique branch condition $\branch$ is mapped to a fresh boolean variable
$z$. Similarly, $!\branch$ is encoded as $\lnot z$, and $\branch_1 \lor
\branch_2$ is encoded as $z_1 \lor z_2$. Then to check an implication
$\branch_1 \Rightarrow \branch_2$, \name{} uses a SAT
solver to check the implication of the encoding. While this check could
err in either direction (due to not modeling the semantics of the
$\branch_i$ precisely), we found it works surprisingly well in
practice. In case the implication check fails due to lack of
precision, we fall back on the original $\oplus$ form which represents
the complete program $\eif{\branch_1}{\expr_1}{\texttt{\textbf{if}
$\ldots$}}$ without loss of precision. Should the implication check
incorrectly succeed, it will be caught by running the merged program against the assertions.

\paragraph*{Constructing the Final Program.}

\begin{algorithm}[t]
\caption{Merge programs}
\label{alg:merge-prog}
\begin{algorithmic}[1]
\Procedure{MergeProgram}{candidates = $\{ \langle \expr_i, \branch_i, \specs_i
\rangle \}$}
\State{merged $\gets \{\bigoplus \langle \expr_i, \branch_i, \specs_i
\rangle \}$}\label{line:merge-all-possible}
  \State{$\textrm{final} \gets$ \{\}}
  \ForAll{$m \in$ merged}
\State{$m \gets$ apply (\ref{eq:merge-rule2})-(\ref{eq:merge-rule4}) to
$m$ until no rewrites possible}
\State{$\textrm{final} \gets \textrm{final} \cup \{ m \}$ if $\forall
\langle \precond_i, \postcond_i \rangle \in \textsc{Specs}(m).$}
\State{\qquad\qquad\qquad$\bigwedge\limits_i \textsc{Prog}(m) \vdash
\precond_i; \postcond_i \Downarrow \val$}
  \EndFor
  \State \Return{$\textsc{Prog}(m)$ s.t. $m \in$ final}
\EndProcedure
\end{algorithmic}
\end{algorithm}

Finally, notice that the merge operation $\oplus$ is not associative,
and it may yield different results depending on the order in which it is
applied. Thus, to get the best solution, \name{} uses
Algorithm~\ref{alg:merge-prog}. It builds the
set of all possible merged fragments
(line~\ref{line:merge-all-possible}). Then it simplifies each candidate
solution using the rewrite rules and only considers a candidate valid if it
passes all tests. It returns any such program as the solution.
This branch merging strategy tries all combinations, so it is less sensitive to spec
order than other component based synthesis approaches~\citep{perelman-tds}.
In practice, we found that reordering the specs does not have much
effect.

\subsection{Discussion}

Before discussing our implementation in the next section, we briefly
discuss some design choices in our algorithm.


Our effect system uses pairs of read and write effects in
regions. As mentioned, this core idea could be extended to any effects 
in a test assertion that can be paired with an effect in the
synthesized method body. For example, throwing and catching exceptions, I/O to disk
or network, or enabling/disabling features in a UI could all be
expressed this way. We leave exploring such effect pairs to future
work.


One convenient feature of our algorithm is that correctness is
determined by passing specs, which are directly executed. Thus, the synthesizer can generate as
many candidates as it likes---i.e., be as over approximate as it
likes---as long as its set of candidates includes the solution. This
feature enables \name{} to use a fairly simple effect annotation
system compared to effect analysis tools~\citep{bocchino-rpl}.

We could potentially adapt our algorithm to work in a capability-based setting, using the observation that capabilities and effects are related~\citep{Craig2018,Brachthauser2020,Gordon2020}. In this setting, assertion failures in tests would indicate specific capabilities needed by the synthesized code. We leave exploring this idea further to future work.

Finally, we distinguish typed holes from effect holes, rather than have
a single type-and-effect hole, to control where to use type-guidance and
where to use effect-guidance. When initially trying to synthesize a
method body, we omit effects because it is unclear which effects are
needed. For example, in Figure~\ref{fig:overview-spec}, the second spec
has read effects on all fields of the post, and yet the target method
does not write any fields, as the spec is checking the case when the
post is not modified. Thus, we cannot simply compute the union of all
read effects in all assertions and use those for effect guidance.
Moreover, type-guided synthesis often
will synthesize effectful expressions, e.g., the call to
\code{Post.where} in Figure~\ref{fig:overview-diagram-merged-prog}.
Conversely, our algorithm only places effect holes in positions where
the type does not matter---hence type information for such a hole would
not add anything. Nonetheless, type-and-effect holes would be a simple
extension of our approach, and we leave exploration of them to future
work in other synthesis domains.

\section{Implementation}
\label{sec:implementation}

\name{} is implemented in approximately 3,600 lines of Ruby, excluding
its dependencies.



Synthesis specifications, as discussed in \S~\ref{sec:overview}, are
written in a custom domain-specific language. Each has
the form:
\begin{lstlisting}
define :name, "method-sig", [consts,...] do
  spec "spec1" do setup { ... } postcond { ... } end ...
end
\end{lstlisting}
where \code{:name} names the method to be synthesized;
\code{method-sig} is its type signature; and \code{consts}
lists constants that can be used in the synthesized method. Each \code{spec} is a
test case the method must pass: \code{setup}
describes the test case setup, and \code{postcond} makes assertions about
the results.

In Ruby, \code{do...end} and \code{\{...\}} are equivalent syntax for
creating \emph{code blocks}, i.e., closures. Having the setup and
postcondition in separate code blocks allows \name to run the setup code
and check the postcondition independently.



\name also has optional hooks for resetting the global state
before any \code{setup} block is run. This ensures candidate
programs are tested in a clean slate without being affected by
side-effects from previous runs. In our experiments, \name resets the
global state by clearing the database.

\paragraph*{Program Exploration Order.}

While our synthesis rules are non-deterministic, our
implementation is completely deterministic. This makes it sensitive to
the order in which expressions are explored. \name uses two metrics to
prioritize search. First, programs are explored
in order of their size; smaller programs are preferred over larger ones.
Program size is calculated as the number of AST nodes in the program.

Second, \name{} prefers trying effect-guided synthesis for expressions
that have passed more assertions rather than fewer.
(Appendix~\ref{subsec:evaluation-rules} formally describes counting
passed assertions.) Untested candidates are assumed to have passed zero
assertions. In general, expressions are explored in decreasing order of
number of passed assertions, then in increasing order of program size.

These metrics combined also help when \name{} synthesizes a candidate
that does not make any progress towards a solution: after running tests and
effect-guided synthesis on such candidates, their size increases, but if
they do not pass more assertions, they are pushed further down the
search queue. We leave experimenting with other search strategies to
future work.





\paragraph*{Effect Annotations.}

We extended RDL to support effect annotations along with type annotations for
library methods. Programmers specify read and write effects following
the grammar in \S~\ref{sec:formalism}. For example a method
annotated with a write effect \code{Post.author} writes to some region
\code{author} in some object of class \code{Post}. Here \code{author}
is an uninterpreted string, selected by the programmer. Similarly the labels
``$.$'' and ``$*$'' stand for pure and any region (or simply ``impure''), respectively. A
region \code{Post.*} is written as \code{Post} for
convenience. One important
extension is a \code{self} effect region, which indicates a read or
write to the class of the receiver. This is essential for supporting
ActiveRecord, whose query methods are inherited by the actual
Rails model classes. For example, we use the \code{self} effect on the
\code{exists?} query method of \code{ActiveRecord::Base}. Then at a call
\code{Post.exists?}, where \code{Post} inherits from
\code{ActiveRecord::Base}, we know the query reads the \code{Post}
table and not any other table.

Effect annotations are similar to frame conditions
\cite{borgida1995frame, meyer2015framing,filliatre2013why3} used in
verification literature. More precise effect annotations help \name
find a solution faster because it will have fewer methods with
subsumed effects than an imprecise one, shrinking the search
space. But effect precision does not affect the correctness of the
synthesized program, since correctness is ensured by the specs. For
example, if the effect annotation for the method {\sf
  Post\#title=} shown in \S~\ref{subsec:overview-type-directed} had
just \texttt{Post} as its write annotation, synthesis would still
work, but would try more candidate programs. In some cases, coarse
effects are required, e.g. the {\sf Post.where} method queries records
from the {\sf Post} table.  It has the coarser {\sf Post} annotation
because which columns such a query will access cannot be statically
specified: it depends on the arguments.
We evaluate some of the tradeoffs in effect precision in \S~\ref{subsec:effect-precision-expt}.

\paragraph*{Type Level Computations.}

\name{} uses RDL~\citep{rdl-github,ren-rdl} to reason
about types, e.g., checking if one type is a subtype of another, and
using the type environment and class table to find terms that can fill
holes.
RDL includes \emph{type-level
  computations}~\citep{kazerounian-comptype}, or \emph{comp types}, in
which certain methods' types include computations that run during type
checking. For example, a comp type for the \code{ActiveRecord\#joins}
method can compute that \code{A.joins(B)} returns a model that
includes all columns of tables \code{A} and \code{B} combined. Using a comp
type for \code{joins} encodes a quadratic number of type signatures,
for different combinations of receivers and arguments, into a single
type, and more for joins of more than two
tables~\citep{kazerounian-comptype}.

\name{} uses RDL's comp types, but with new type signatures
designed for synthesis. In particular, the previous version of RDL's
comp types gave precise types when the receiver and arguments were
known, e.g., in \code{A.joins(B)}, RDL knows exactly which two classes
are being joined. But this may not hold during synthesis, e.g., if
\code{B} is replaced by a hole in the example, then  the exact return
type of the \code{joins} call cannot be computed.

To address this issue, we modified RDL's existing comp type signatures
for \code{ActiveRecord} methods like \code{joins} so that they compute
all possible types. For example, if a hole is an argument to
\code{joins}, then the type finds all models \code{B1}, \code{B2},
$\ldots$ that could be joined (i.e., those with associations); gives
the hole type $\code{B1} \cup \code{B2} \cup \ldots$; and sets the
return type of joins to a table containing the columns of $\code{A},
\code{B1}, \code{B2}, \ldots$. This over-approximation is narrowed as the
argument terms are synthesized, leading to cascading narrowing of types
throughout the program as discussed in \S~\ref{subsec:type-directed}.

\paragraph*{Optimizations.}

Synthesis of terms that pass a spec is an expensive procedure.
In practice, we found solutions to a single spec often satisfy others.
Thus, when confronted with a new spec, \name{} first tries
existing solutions and conditionals to see if they hold for the spec,
before falling back on synthesis from scratch if needed. This makes
the bottleneck for synthesis not the number of tests, but the number of
unique paths through the program. Moreover, this reduces the number of tuples for merging, as a single expression and conditional tuple can represent
multiple specs $\specs$.

Finally, we found that in practice, the
condition in one spec often turns out to be the negation of the condition in another. Thus during synthesis of conditionals, \name tries
the negation of already synthesized conditionals before falling
back on synthesis from scratch.

\paragraph*{Limitations.}

While \name{} works on a wide range of programs, as we will demonstrate next, it does have several key limitations. First, \name{} currently only synthesizes code that does not need type casts to be well-typed. This ensures programs do not have type errors at run time, but eliminates some valid programs from consideration. Second, the set of constants \name{} can use during synthesis is fixed ahead of time. This places programs that use unlikely constants out of reach, e.g., we have encountered Rails model methods that include raw SQL query strings (instead of only using ActiveRecord). Finally, because \name{} uses enumerative search, it can face a combinatorial explosion when searching for nested method calls, e.g., if there are $n$ possible method calls, available, synthesizing 
\emethcall{\class}{\meth}{\emethcall{\class}{\meth}{\emethcall{\class}{\meth}{\var}}} may require an $O(n^3)$ search.
In practice, we did not face this problem as deeply nested method
calls are rarely used in Rails apps.

\section{Evaluation}
\label{sec:evaluation}

We evaluated \name by using it to synthesize a range of benchmarks extracted from widely used open source applications that use a variety of libraries. We pose the following questions in our evaluation:

\begin{itemize}
  \item How does \name{} perform using code based on existing unit tests in widely deployed applications? (\S~\ref{subsec:syn-bench-results})
  \item How much improvement is type-and-effect guidance compared to alternatives such as only type-guidance or only effect-guidance? (\S~\ref{subsec:syn-comparison-results})
  \item How does the precision of effect annotations affect synthesis performance? (\S~\ref{subsec:effect-precision-expt})
\end{itemize}


\subsection{Benchmarks}

To answer the questions above, we collected a benchmark suite
comprised of programs from the following sources:

\begin{itemize}
  \item \emph{Synthetic benchmarks} is a set of minimal examples that
  demonstrate features of \name.
  \item \emph{Discourse} \citep{discourse} is a
  Rails-based discussion platform used by over 1,500 companies and
  online communities.
  \item \emph{Gitlab} \citep{gitlab} is a web-based Git repository
  manager with wiki, issue tracking, and CI/CD tools built on Rails.
  \item \emph{Diaspora} \citep{diaspora} is a distributed
  social network, with groups of independent nodes (called Pods), also built
  on Rails.
\end{itemize}

We selected these apps because they are popular, well-maintained,
widely used, and representative of programs that are written with
supporting unit tests.
We selected a subset of the app's methods for synthesis, choosing ones
that fall into the Ruby grammar we can synthesize: method calls, hashes,
sequences of statements and branches. We currently do not synthesize blocks
(lambdas), for/while loops, case statements, or meta-programming in the
synthesized code. All benchmarks from apps have side effects due to
either database accesses or reading and writing globals.

\begin{table*}
\centering
\small
\begin{tabular}{|r|r|r|r|r|r|r|r||r|r|r|r||r|r|}
  \hline
  \thead{\multirow{2}{*}{\rotatebox[origin=c]{90}{Group}}} & & & \thead{\#} & \multicolumn{2}{c|}{\thead{Asserts}} & \thead{\# Orig} & \thead{\# Lib} & \multicolumn{4}{c||}{\thead{Time (sec)}} & \thead{Meth} & \thead{\# Syn} \\
  \cline{5-6} \cline{9-12}
    & \thead{ID} & \thead{Name} & \thead{Specs} &
  \thead{Min} & \thead{Max} & \thead{Paths} & \thead{Meth} & \thead{Median $\pm$ SIQR} & \thead{Types} &
  \thead{Effects} &
  \thead{Neither} & \thead{Size} & 
  \thead{Paths} \\  
  \hline
  \multirow{7}{*}{\rotatebox[origin=c]{90}{Synthetic}}
  & S1 & lvar             &     1 &        1 & 1 & 1 & 164 &  \btime{0.34}{0.01} &   1.36 & 11.97 & - &  4 &        1 \\
  & S2 & false            &     1 &        1 & 1 & 1 & 164 &  \btime{0.35}{0.01} &   1.37 & 12.19 & - &  4 &        1 \\
  & S3 & method chains    &     2 &        1 & 1 & 1 & 164 &  \btime{0.98}{0.01} &   9.56 &     - & - & 10 &        1 \\
  & S4 & user exists      &     2 &        1 & 1 & 1 & 164 &  \btime{0.98}{0.02} &   9.52 &     - & - &  9 &        1 \\
  & S5 & branching        &     3 &        1 & 1 & 2 & 165 &  \btime{2.49}{0.07} &  38.37 &     - & - & 17 &        2 \\
  & S6 & overview (ext)   &     3 &        4 & 4 & 3 & 164 & \btime{12.78}{0.09} &      - &     - & - & 72 &        3 \\
  & S7 & fold branches    &     3 &        1 & 1 & 1 & 164 & \btime{82.44}{0.95} & 218.51 &     - & - & 13 &        1 \\
  \hline
  \multirow{4}{*}{\rotatebox[origin=c]{90}{Discourse}}
  & A1 & User\#clear\_glob\ldots &     3 &        2 & 2 & 3 & 169 &  \btime{2.11}{0.04} & - & - & - &   24 &        3 \\
  & A2 & User\#activate          & 2 (3) &        1 & 4 & 2 & 170 &  \btime{8.95}{0.23} & - & - & - &   28 &        2 \\
  & A3 & User\#unstage           & 3 (4) &        1 & 5 & 2 & 164 & \btime{50.02}{0.55} & - & - & - &   31 &        2 \\
  & A4 & User\#check\_site\ldots &     5 &        1 & 1 & 2 & 168 & \btime{51.6}{0.23} & - & - & - &   28 &        3 \\
  \hline
  \multirow{4}{*}{\rotatebox[origin=c]{90}{Gitlab}}
  & A5 & Discussion\#build        &     1 &         4 & 4  & 1 & 167 & \btime{0.24}{0.01} &     - &    - &     - & 18 &        1 \\
  & A6 & User\#disable\_two\ldots &     1 &        10 & 10 & 1 & 164 & \btime{0.25}{0.01} &     - & 0.44 &     - & 22 &        1 \\
  & A7 & Issue\#close             & 1 (2) &         3 & 3  & 1 & 166 & \btime{0.77}{0.03} & 25.99 & 0.13 &  0.37 & 15 &        1 \\
  & A8 & Issue\#reopen            & 1 (3) &         5 & 5  & 1 & 166 & \btime{3.68}{0.1} &     - & 0.55 & 45.66 & 17 &        1 \\
  \hline
  \multirow{4}{*}{\rotatebox[origin=c]{90}{Diaspora}}
  &  A9 & Pod\#schedule\_\ldots     & 3 (4) &        1 & 1 & 2 & 161 & \btime{2.44}{0.04} &    - & - &    - &  19 &        2 \\
  & A10 & User\#process\_inv\ldots  &     1 &        2 & 2 & 2 & 165 & \btime{2.64}{0.05} & 0.81 & - & 0.85 &  12 &        1 \\
  & A11 & InvitationCode\#use!      &     1 &        1 & 1 & 1 & 165 & \btime{4.23}{0.06} &    - & - &    - &  12 &        1 \\
  & A12 & User\#confirm\_email      &     7 &        4 & 4 & 2 & 166 & \btime{7.28}{0.11} &    - & - &    - &  31 &        3 \\
  \hline
\end{tabular}
\caption{Synthesis benchmarks and results. \emph{\# Specs} is the number
of specs used to synthesize the method; \emph{Asserts} reports the minimum and maximum number of assertions over all
specs for every benchmark; \emph{\# Orig Paths} is the number of
paths through the method as written in the app; \emph{\# Lib Meth} is the
number of library methods used for every benchmark; \emph{Time} shows the
median and semi-interquartile range over 11 runs, followed by the median
time for synthesis using only types, only effects and naive term
enumeration (\emph{Neither}). \emph{Meth Size}
is the number of AST nodes in the synthesized method; \emph{\# Syn Paths}
shows the number of paths through the synthesized method.}
\label{table:results}
\end{table*}

Table~\ref{table:results} lists the benchmarks. The first column group
lists the app name (or \emph{Synthetic} for the
synthetic benchmarks); the benchmark id; the benchmark name; and the
number of specs. The synthetic benchmarks exercise features of \name{}
by synthesizing pure methods, methods with side effects, methods in
which multiple branches are folded into a single line program, etc. The
Discourse benchmarks include a number of effectful methods in the {\sf
User} model, such as methods to activate an user account,
unstage a placeholder account created for email integration, etc.
The Gitlab benchmarks include methods that disable two factor
authentication for a user, methods to close and reopen
issues, etc. Finally, the Disaspora benchmarks include methods to
confirm a user's email, accept a user invitation, etc.

We derived the specs for the non-synthetic benchmarks directly from the
unit tests included in the app. We split each test into \emph{setup} and
\emph{postcondition} blocks in the obvious way, and
we added an appropriate type annotation to the
synthesis goal.
Across all benchmarks, we started with a base set of constants ($\Sigma$
in \S~ \ref{sec:formalism}) to be \vtrue, \vfalse, 0, 1 and the empty
string. Then we added
\vnil and singleton classes (for calling class methods) on
a per benchmark basis as needed.
(As with many enumerative search based methods, we rely on the user to provide the right set of constants.)

A few apps have several different unit tests with exactly the same setup
but different assertions in the postcondition. We merged any such group
of tests into a single spec with that setup and the union of the
assertions as the postcondition, to ensure that every spec setup can be
distinguished with a unique branch condition, if necessary. We indicate
this in the \emph{\# Specs} column of Table~\ref{table:results} by listing
the final number of specs followed by the original number of tests in
parentheses if they differ. We report the minimum and maximum number of
assertions over all specs per benchmark in the \emph{Asserts} columns
and the number of paths through the method in the true canonical
solution (from the app) in the \emph{\# Orig Paths} column.

\paragraph*{Annotations for Benchmarks.}

Finally, the \emph{\# Lib Meth} column lists the number of library
methods available during synthesis. These are methods for which we
provided type-and-effect annotations. In total, 164 such methods are
shared across all benchmarks, including, e.g., ActiveRecord and core
Ruby libraries. Since our benchmarks are sourced from full apps, they
often also depend on some other methods in the app. We wrote
type-and-effect annotations for such methods and included those
annotations only when synthesizing that app. Since \name{} needs to run
the synthesized code, when running specs we include the code for both
general-purpose methods, such as those from ActiveRecord, and required
app-specific methods. We slightly modify the set of library methods for
A9, as discussed further below.



To find effect labels for app-specific methods, we found examining the method name and quickly scanning its code was typically quite helpful. Often it was clear if a method was pure or impure. For impure methods, there were a few cases. Sometimes, methods access the same object fields irrespective of how the method is called, so we give such methods the most precise labels, e.g., we used effect \code{InvitationCode.count} for benchmark A10. Other times, it is apparent the method accesses different fields of a class depending on the method's arguments or the global state, so we give these class effect labels, e.g., \code{User} (equivalent to \code{User.*}). Overall, the simplicity of the effect system helped here, as we could use human-readable region identifiers even without any object references, e.g., the effect \code{InvitationCode.count} abstracts over all possible instances of \code{InvitationCode} class.

The other main category of effect labels was for Rails libraries such as ActiveRecord. We constructed these labels by following the documentation. For metaprogramming-generated column accessor methods, we extended RDL's existing type generating annotations~\cite{ren-rdl} to also generate effects. For example, when RDL creates the type signature for an accessor method \code{Post\#title} for the {\sf title} column of the {\sf Post} table, it now also creates a read effect annotation \code{Post.title} for it.

Overall, we found writing effect annotations to be easier than our previous efforts writing type annotations for Ruby~\cite{ren-rdl,kazerounian-comptype}, though of course we relied on that previous experience. We leave a systematic evaluation of the effort of writing effect annotations to future work.


\subsection{Synthesis Correctness and Performance}
\label{subsec:syn-bench-results}

\name{} successfully synthesized methods that pass the specs for every benchmark. We manually examined the output and found that the synthesized code is equivalent to the original, human-written code, modulo minor differences that do not change the code's behavior in practice. For example, one such difference occurs with original code that updates multiple database columns with a single ActiveRecord call, and then has a sequence of asserts to check that each updated column is correct. Because \name{} considers the effects of assertions in the postcondition one by one, it instead synthesizes a sequence of database updates, one per column. Another difference occurs in Gitlab, which uses the {\sf state\_machine} gem (an external package) to maintain an issue's state (closed, reopened, etc). \name synthesizes correct implementations that work without the gem.


The middle group of columns in Table~\ref{table:results} summarizes
\name{}'s running time. We set a timeout of 300 seconds on all
experiments. The first column reports performance numbers for the full
system as the median and semi-inter\-quar\-tile range (SIQR) of 11 runs
on a 2016 Macbook Pro with a 2.7GHz Intel Core i7 processor and 16GB
RAM. The next three columns show the median performance when \name uses
only type-guidance, only effect-guidance, and naive enumeration,
respectively. The SIQRs (omitted due to space constraints) for these runs are very small compared to the
median runtime, similar to the performance numbers with all features
enabled. We discuss the runs with certain guidance disabled
in detail in \S~\ref{subsec:syn-comparison-results}. The
right-most group of columns shows the synthesized method size (in terms
of number of AST nodes) and the number of paths through the method (1 for
straight-line code).



Overall, \name{} runs quickly, with around 80\% of benchmarks solving in
less than 9s. Benchmarks like A3 take longer because it requires
synthesis of \vnil terms---recall \vnil is the bottom element of our
type lattice, causing \name{} to synthesize \vnil at every typed hole
for method arguments.
Consequently, this requires testing all
completed candidates---even though they eventually fail---consuming
significant time.

For one benchmark, A9, we changed the set of default library methods slightly due to some pathological behavior. This benchmark includes an assertion that invokes ActiveRecord's \code{reload} method, which touches all fields of that record. But then when \name{} tries to find matching write effects, it explores a combinatorial explosion of writes to different subsets of the fields. This effort is almost entirely wasted, because the remainder of the assertion looks at only one particular field---but that one read is subsumed by the effect of the \code{reload}, making it invisible to \name{}'s search. As a result, synthesis for A9 slows down by two orders of magnitude. We addressed this by removing four ActiveRecord methods that manipulate specific fields and adding ActiveRecord's \code{update!} method as the only way to write a field back to the database. An alternative approach would have been to move the \code{reload} call to be outside the assertion.

As this example shows, and as is common with many synthesis problems, performance is very hard to predict. Indeed, we can see from Table~\ref{table:results} that performance is generally not well
correlated with either the size of the output program or with the number
of branches. The number of assertions (which direct the side effect
guided synthesis) does not correlate with the synthesis time. We do observe that \name's branch merging strategy is
effective, often producing fewer conditionals than there are specs, e.g.,
in A12 there are seven specs but only three conditionals. Though,
sometimes the results are not always optimal if the branch
merging strategy finds a program that passes all tests, but a program
with fewer branches exists, e.g., for A4 and A12,
\name{} produces a program with one more branch than the hand-written.




\subsection{Performance of Type- and Effect-Guidance}
\label{subsec:syn-comparison-results}

\begin{figure}
\centering
\includegraphics[scale=0.48]{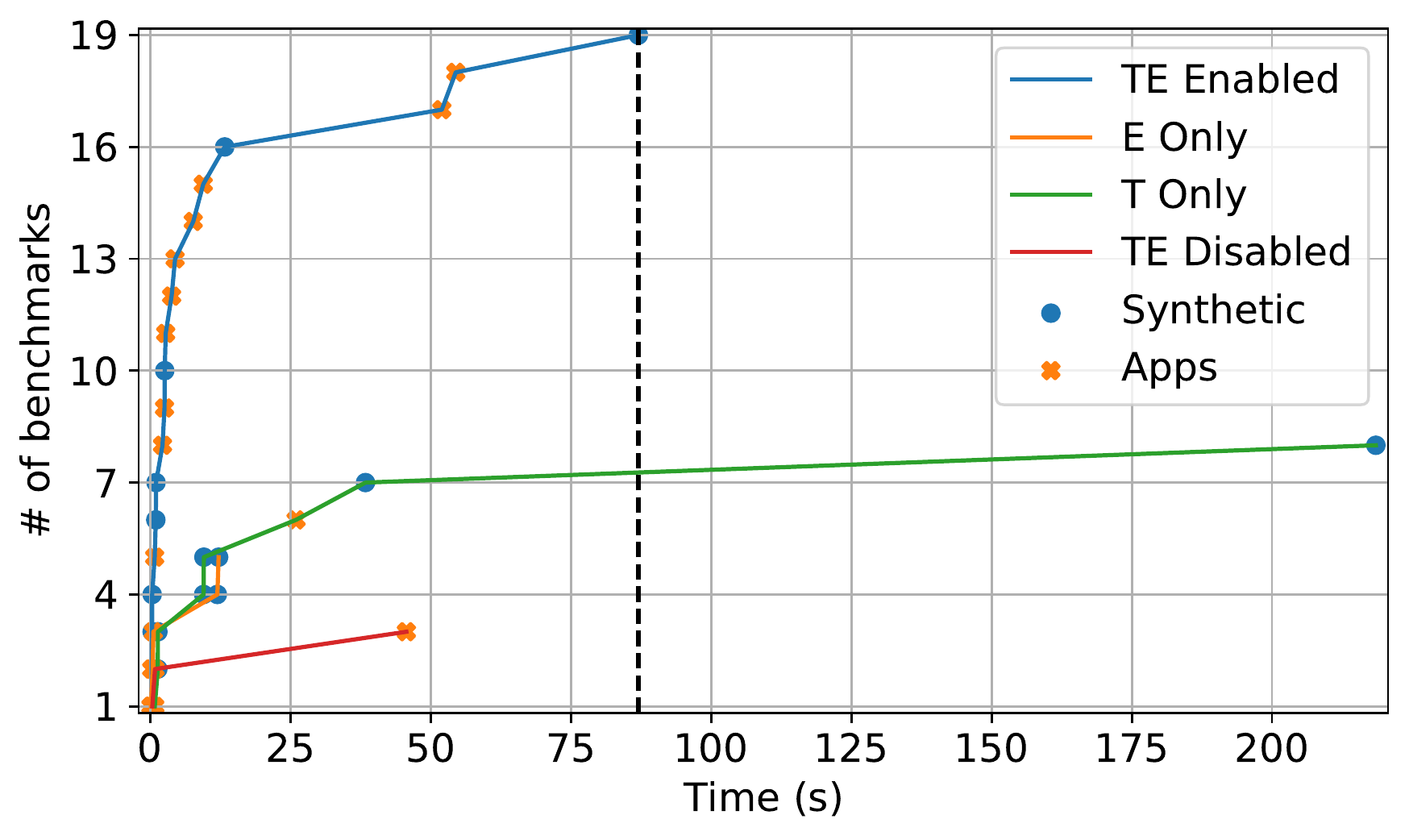}
\caption{Number of benchmarks synthesized using type-and-effect
(\emph{TE Enabled}) guided synthesis relative to using only type
(\emph{T Only}) or effect (\emph{E Only}) guidance separately and naive
enumeration (\emph{TE Disabled}). Higher is better.}
\label{fig:rbsyn-comparison}
\end{figure}

Next, we explore the performance benefits of type- and effect-guidance.
Figure~\ref{fig:rbsyn-comparison} plots the running times from
Table~\ref{table:results} when all features of \name are enabled
(\emph{TE Enabled}), with only type-guidance (\emph{T Only}), with only
effect-guidance (\emph{E Only}) and with neither (\emph{TE Disabled}).
The plot shows the number of benchmarks that complete ($y$-axis) in a
given amount of time ($x$-axis), based on the median running times.
This experiment serves as a proxy to show how a synthesis procedure that uses type-guidance but not effect-guidance, such as \textsc{SyPet}~\cite{feng-componentsyn} or
\textsc{Myth}~\cite{osera2015type,frankle2016example}, may have performed if adapted for Ruby.


We can clearly see that type- and effect-guided synthesis performs best,
successfully synthesizing all benchmarks; the slowest takes 83s. In
contrast, with both strategies disabled, all but three small benchmarks
time out. Performance with only type- or only effect-guidance lies in
between. With only type-guidance, synthesis completes on eight
benchmarks, of which the majority are pure methods from the synthetic
benchmarks. From apps, it only synthesize A7 and A10. In these benchmarks, the needed effectful expressions are small and
hence can be found with essentially brute-force search. With only
effect-guidance, synthesis performance significantly worse, completing
only five benchmarks, of which only three are from apps. These benchmarks succeeded because effect-guided synthesis
quickly generates the template for the effectful method calls and then
correctly fills them since they are small and can be found quickly by naive enumeration.

\subsection{Effect Annotation Precision vs. Performance}
\label{subsec:effect-precision-expt}

\begin{figure}
\centering
\includegraphics[scale=0.42]{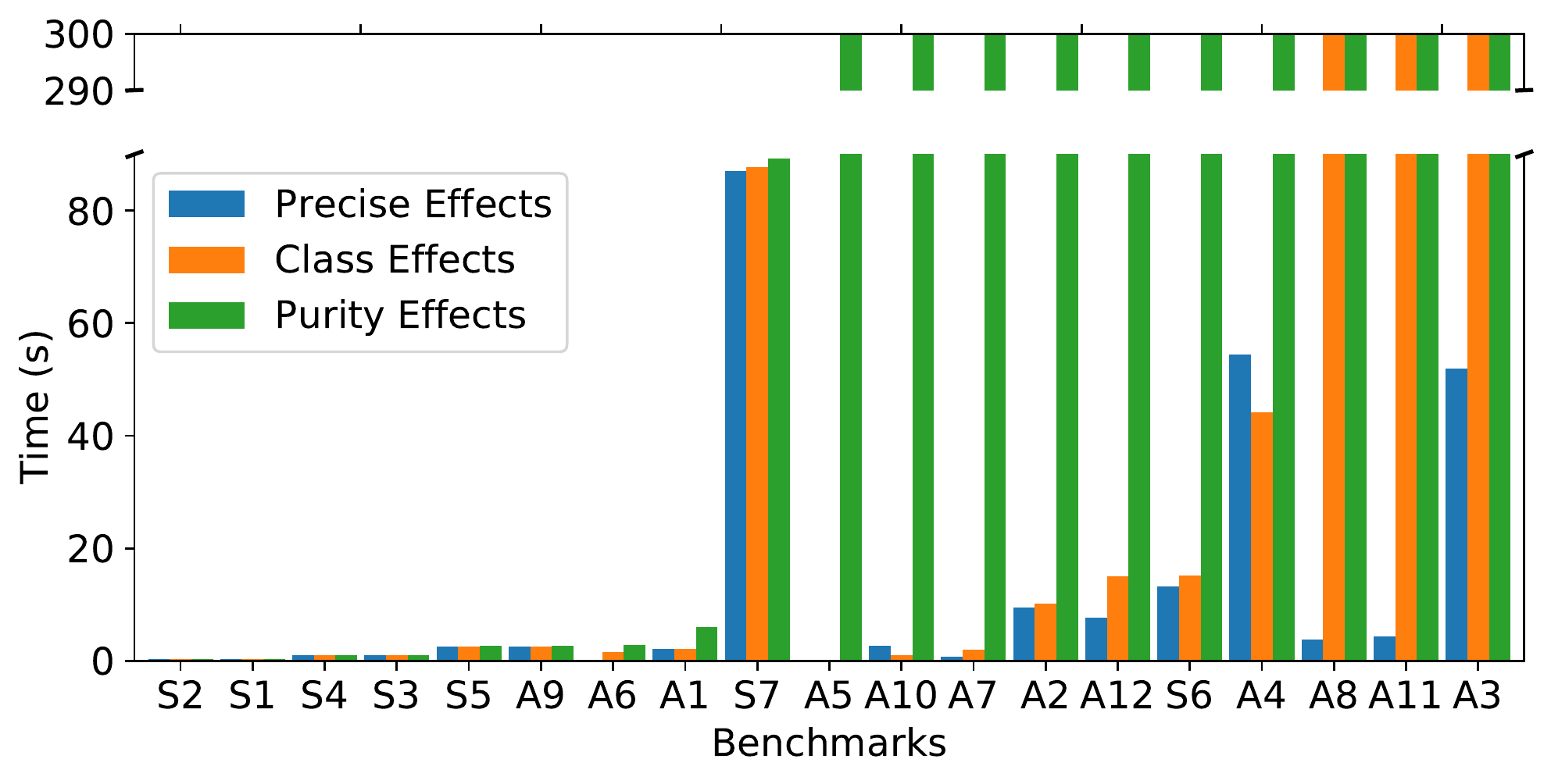}
\caption{Performance of \name with varying effect annotation precision:
full, class effects only, and
purity annotations on library methods. Lower is better. Full height indicates timeout.}
\label{fig:rbsyn-effect-precision}
\end{figure}

Finally, we explore the tradeoff between effect annotation precision and
synthesis performance. Recall that we found writing effect
annotations easier for our benchmarks than writing type
annotations. However, the effort can be further minimized by writing less
precise annotations. This will not affect correctness, since \name{}
only accepts synthesis candidates that pass all specs, but it does
affect performance. 

Figure~\ref{fig:rbsyn-effect-precision} plots the median of synthesis
times for benchmarks over 11 runs under three conditions:
\emph{Precise Effects}, which are the effects used above; \emph{Class Effects},
in which annotations include only class names and eliminate region
labels (e.g., \code{Post.title} becomes \code{Post}); and \emph{Purity
Effects}, in which the only effect annotations are pure or impure (the $\pure$ and
$*$ effects, respectively, in our formalism). The benchmarks (\emph{x}-axis) are ordered
in increasing order of time for \emph{Purity Effects}, then
\emph{Class Effects}, and finally \emph{Precise Effects}.

From these experiments, we see that synthesis time increases as
effect annotation precision decreases, often leading to a timeout. Class
labels were sufficient to synthesize 16 of 19 benchmarks.
Overall, class labels take
time similar to precise labels, except for the three cases (A8, A11, and
A3) where side-effecting method calls require precise labels to quickly
find the candidate. As all precise effects are reduced to class effects,
\name must try many candidates with class effect before finding
the correct one, leading to timeouts.

We note that A1 and A4 are slightly faster when using class effects. The reason is an implementation detail. The effect holes in these benchmarks can only be correctly filled by methods whose regular annotations are class annotations (more precise annotations are not possible). However, when trying to fill holes, \name{} first tries all methods with precise  annotations, only afterward trying methods with class annotations. Since the precise annotations never match, this yields worse performance under the precise effect condition than under the class effect condition,  when the search could by chance find the matching methods sooner.

Purity labels only enabled synthesis of 9
benchmarks, including just 3 of 12 app benchmarks. The purity
annotations are slow in general and only effective in the cases where
the number of impure library methods is small.



\section{Related Work}

\paragraph*{Component-Based Synthesis.}

Several researchers have proposed component-based synthesis, which
creates code by composing calls to existing APIs, as \name{} does. For example, \citet{jha2010oracle} propose
synthesis of loop-free programs for bit-vector
manipulation. Their approach uses formal
specifications for synthesis, in contrast to \name, which uses unit tests.
\textsc{Hoogle+}~\cite{james2020digging} uses
Haskell tests and types to synthesize potential solutions, primarily geared towards
API discovery. \textsc{CodeHint}~
\citep{galenson2014codehint} synthesizes Java programs, using a
probabilistic model to guide the search towards expressions more often
used in practice. \textsc{SyPet}~\citep{feng-componentsyn}
also synthesizes programs that use Java APIs, by modeling them as a
petri net and using SAT-based techniques to find a solution. These approaches do
not support synthesis of programs with branches, which are common in the
domain of web apps. While \textsc{SyPet} supports synthesis with
side-effecting methods and \textsc{CodeHint} detects undesirable side
effects during the search and avoids them, \name uses side effect
information from test cases to guide the search.

\paragraph{Programming by Example.}

\textsc{Myth}~\citep{osera2015type,frankle2016example} uses
bidirectional type checking to synthesize programs, using input/output
examples as the specification. However, \textsc{Myth} expects examples to
be \emph{trace complete}, meaning the user has to provide input/output
examples for any recursive calls on the function arguments. \name does not synthesize
recursive functions, as they are rarely needed in our target domain of Ruby web apps.
\textsc{Escher}~\citep{albarghouthi2013recursive} and spreadsheet
manipulation tools~
\citep{gulwani2011automating,harris2011spreadsheet,gulwani2012spreadsheet}
all accept input/output examples as a partial specification for
synthesis. These tools primarily target users who cannot
program, whereas \name is targeted towards programmers. In addition, \name's specs are full unit tests, so they can check both return values and
side effects. $\lambda^2$~\citep{feser2015synthesizing} synthesizes data
structure transformations using higher-order functions, a feature not
handled by \name because of our target domain of Rails web apps, 
which rarely use such functions. STUN~\cite{alur2015synthesis} uses a program merging strategy that is
similar to ours, but it depends on defining domain-specific unification
operators to safely combine programs under branches.
In contrast, our
approach may be more domain-independent, using preconditions and tests to find correct branch conditions.
There have been
multiple approaches to synthesizing database programs 
\citep{cheung2013optimizing,feng2017component}. Perhaps the closest in
purpose to \name is \textsc{Scythe}~\citep{wang2017synthesizing},
which synthesizes SQL queries based on input/output examples.
\textsc{Scythe} uses a two-phased synthesis process to synthesize an abstract query,
after which enumeration is used to concretize the abstract query. In contrast, the
use of comp types~\cite{kazerounian-comptype} allows \name to quickly construct
a template for a database query. With precise types for the
method argument holes, this essentially builds abstract queries for
free, whose holes are then filled later during synthesis.

\paragraph*{Solver-Aided Synthesis.}

In solver-aided synthesis, synthesis specifications are transformed
to a set of constraints for a SAT or SMT solver.
\textsc{Synquid}~\citep{polikarpova-synquid}
uses polymorphic refinement types as the specification for synthesis.
\textsc{Lifty}~\citep{polikarpova2020lifty} is a similar type system that
verifies information flow control policies and synthesizes program repairs as needed to satisfy the policies.
Both \textsc{Synquid} and \textsc{Lifty} synthesize conditionals using logical abduction. In contrast, \name{} uses branch merging to synthesize conditionals, since translating Rails code and libraries into logical formulas is impractical.

Sketch~\citep{solar2006combinatorial} allows
users to write partial programs, called sketches, where the omitted
parts are then synthesized by the tool.
\textsc{Migrator}~\citep{wang2019synthesizing} uses
\emph{conflict-driven learning}~\citep{feng2018program} to
synthesize raw SQL queries, for use in database programs for schema
refactoring. In contrast, programs synthesized by \name{} use
ActiveRecord to access the database.
Rosette~\citep{torlak2014lightweight,torlak2013growing} is a
solver-aided language that provides access to verification and
synthesis. It relies on symbolic execution, and thus
requires significant modeling of external libraries for synthesizing
programs that use such libraries.

\textsc{eusolver}~\citep{alur2017scaling} synthesizes programs with
branches, using an information-gain heuristic via decision tree
learning. While, the decision tree learning procedure can produce
branches in an enumerative search setting (provided the input/output
example set is complete), we leave an exploration of
how it compares to our rule-based merging to future work.
However, \textsc{eusolver} requires a SMT solver to produce counterexamples
to build the input/output example set which has the additional cost of 
requiring formal specifications of library method semantics, an
impractical task in the Rails setting.
SuSLik~\cite{polikarpova2019structuring} synthesizes
heap-manipulating programs using separation logic to precisely model the
the heap. \name, in contrast, uses very coarse effects to track accesses
that can go beyond the heap, such as database reads and writes.

\section{Conclusion}

We presented \name, a system for type- and effect-guided program synthesis for Ruby. In \name{}, the synthesis goal is described by the target method type and a series of specs comprising preconditions followed by postconditions that use assertions. The user also supplies the set of constants the synthesized method can use, and type-and-effect annotations for any library methods it can call. \name{} then searches for a solution starting from a hole $\ehole:\type$ typed with the method's return type, inserting (write) effect holes $\effhole:\eff$ derived from the read effects of failing assertions. Finally, \name{} merges together solutions for individual specs by synthesizing branch conditions to select among the different solutions as needed. We evaluated \name{} by running it on a suite of  19 benchmarks, 12 of which are representative programs from popular open-source Ruby on Rails apps. \name synthesized correct solutions to all benchmarks, completing synthesis of 15 of the 19 benchmarks in under 9s, with the slowest benchmark solving in 83s. We believe \name demonstrates a promising new approach to synthesizing effectful programs.

\begin{acks}                            
  Thanks to the anonymous reviewers for their helpful comments.
  This research was supported in part by NSF CCF-1900563, and NSF
  CCF-1846350.
\end{acks}

\balance

\bibliography{bibfile}

\newpage
\appendix
\section{Appendix}
\label{sec:appendix}

\subsection{Evaluation Rules}
\label{subsec:evaluation-rules}

\begin{figure}
  \centering
  
  $$
  \begin{array}{lccl}
 \emph{Errors} \quad
 & \mathcal{E}
 & ::= & \err{\eff_r}{\eff_w} \\
 \emph{Results} \quad
 & \mathcal{R}
 & ::= & \val \spmid \mathcal{E} \\
  \end{array}
  $$
  \caption{Extended \corelang.}
  \label{fig:extended-lang}
\end{figure}

\begin{figure*}
\centering
\judgementHead{\llbracket E, c, \rwpair{\eff_r}{\eff_w}, \postcond
\rrbracket \hookrightarrow_{\classtable} \llbracket E', c', \rwpair{\eff'_r}
{\eff'_w}, \postcond' \rrbracket}
\begin{mathpar}

\inference{\val \in \{\vtrue, \classobj{\class}\}}
{\llbracket E, c, \rwpair{\eff_r}{\eff_w}, \eassert{\val}
\rrbracket \hookrightarrow_{\classtable} \llbracket E, c + 1,
\rwpair{\eff_r}{\eff_w}, \val
\rrbracket}[\rulename{E-AssertPass}]

\inference{\val \in \{\vfalse, \vnil\}}
{\llbracket E, c, \rwpair{\eff_r}{\eff_w}, \eassert{\val}
\rrbracket \hookrightarrow_{\classtable} \llbracket E, c, \rwpair{\eff_r}
{\eff_w},
\err{\eff_r}{\eff_w}
\rrbracket}[\rulename{E-AssertFail}]

\inference{\llbracket E, c, \rwpair{\eff_r}{\eff_w}, \expr
\rrbracket \hookrightarrow_{\classtable} \llbracket E', c, \rwpair{\eff'_r}
{\eff'_w}, \expr' \rrbracket}
{\llbracket E, c, \rwpair{\eff_r}{\eff_w}, \eassert{\expr} \rrbracket
\hookrightarrow_{\classtable} \llbracket
E', c, \rwpair{\eff'_r}{\eff'_w}, \eassert{\expr'} \rrbracket}[
\rulename{E-AssertStep}]

\inference{
\textrm{type\_of}(\val_r,\val_a) = (\class_r, \class_a) &
\meth: \type_a \xrightarrow[]{\rwpair{\eff_r}{\eff_w}} \type \in
\classtable(\class) \\
\class_r \leq \class & \class_a \leq \type_a &
\textrm{call}(\class.\meth, \val_r, \val_a) = \val
}
{\llbracket E, c, \rwpair{\eff'_r}{\eff'_w}, \emethcall{\val_r}{\meth}{\val_a}
\rrbracket \hookrightarrow_{\classtable} \llbracket E, c, \rwpair{\eff'_r}
{\eff'_w} \cup \rwpair{\eff_r}{\eff_w}, \val
\rrbracket}[\rulename{E-MethCall}]

\inference{
\llbracket E, c, \rwpair{\eff_r}{\eff_w}, \postcond_1 \rrbracket
\hookrightarrow_{\classtable}
\llbracket E, c, \rwpair{\eff'_r}{\eff'_w}, \postcond_1' \rrbracket}
{\llbracket E, c, \rwpair{\eff_r}{\eff_w}, \eseq{\postcond_1}{\postcond_2}
\rrbracket \hookrightarrow_{\classtable}
\llbracket E, c, \rwpair{\eff'_r}{\eff'_w}, \eseq{\postcond_1'}{\postcond_2}
\rrbracket}[\rulename{E-SeqStep}]

\inference{}
{\llbracket E, c, \rwpair{\eff_r}{\eff_w}, \eseq{\val}{\postcond_2} \rrbracket
\hookrightarrow_{\classtable}
\llbracket E, c, \rwpair{\pure}{\pure}, \postcond_2 \rrbracket}[
\rulename{E-SeqVal}]

\inference{}
{\llbracket E, c, \rwpair{\eff_r}{\eff_w}, \eseq{\err{\eff_r}{\eff_w}}
{\postcond_2} \rrbracket \hookrightarrow_{\classtable} \llbracket E, c,
\rwpair{\eff_r}{\eff_w}, \err{\eff_r}{\eff_w} \rrbracket}[\rulename{E-SeqErr}]

\end{mathpar}
\caption{Selected rules for operational semantics of the
postcondition \postcond}
\label{fig:assert-op-semantics}
\end{figure*}

We extend \corelang to include errors $\mathcal{E}$. Errors can
originate from the evaluation of an assertion \eassert{\expr} and
encapsulates the read effect $\eff_r$ and write effect $\eff_w$ inferred
from \expr. Results $\mathcal{R}$ of an evaluation can either be a value
or an error. Collecting effects while evaluating the postcondition of tests require special evaluation
rules. Figure \ref{fig:assert-op-semantics} shows selected rules of the small step
operational semantics for only postconditions (rest omitted as they are
standard rules). The rules prove judgments of the
form $\llbracket E, c, \rwpair{\eff_r}{\eff_w}, \postcond \rrbracket
\hookrightarrow_{\classtable} \llbracket E', c', \rwpair{\eff'_r}{\eff'_w},
\postcond' \rrbracket$ that reduce configurations that
contain a dynamic environment $E$, counter of passed assertions $c$, the
pair of read and write effects \rwpair{\eff_r}{\eff_w} collected
during evaluation, and postcondition \postcond under evaluation.
Rule \rulename{E-AssertPass} applies when assertion evaluates to a
truthy-y value. It also increments the counter for passed
assertions.
If \expr evaluates to a false-y value, it results in an error,
in which case it returns the collected side effects with the error (
\rulename{E-AssertFail}). Evaluation of a library method, gives a union
of its effects with the already collected effects
(\rulename{E-MethCall}). During
the evaluation of a sequence \eseq{\postcond}{\postcond}, if the first assertion
evaluates to a value, the evaluation continues discarding all
collected effects (\rulename{E-SeqVal}). If the evaluation of an assert yields
an error, the evaluation of postcondition terminates with the error as
final result (\rulename{E-SeqErr}). We define the big step semantics as follows:
$\postcond \Downarrow \mathcal{R}
\ \textnormal{if}\ \exists E', c, \rwpair{\eff_r}{\eff_w}. \llbracket
[\var_r \rightarrow \val], 0,
\rwpair{\pure}{\pure},
\postcond \rrbracket \hookrightarrow^*_{\classtable} \llbracket E',
c, \rwpair{\eff_r}{\eff_w}, \mathcal{R} \rrbracket$, in other words,
evaluating the postcondition in an environment containing the return
value of synthesis goal $\var_r$, will evaluate to a result.

\subsection{Type-Guided Synthesis}
\label{subsec:appendix-type-syn}

\begin{figure*}
  \centering
  \judgementHead{\Sigma, \tenv \vdash_{\classtable}
  \expr \rightsquigarrow \expr: \type}
  \begin{mathpar}
  
  \inference{}{\Sigma, \tenv \vdash_{\classtable}
  \vnil \rightsquigarrow \vnil: \emph{Nil}}[\rulename{T-Nil}]

  \inference{}{\Sigma, \tenv \vdash_{\classtable}
  \vtrue \rightsquigarrow \vtrue: \emph{Bool}}[\rulename{T-True}]
  
  \inference{}{\Sigma, \tenv \vdash_{\classtable}
  \vfalse \rightsquigarrow \vfalse: \emph{Bool}}[\rulename{T-False}]

  \inference{}{\Sigma, \tenv \vdash_{\classtable}
  \classobj{\class} \rightsquigarrow \classobj{\class}: \class}[
  \rulename{T-Obj}]
  
  \inference{}{\Sigma, \tenv \vdash_{\classtable}
  \err{\eff_r}{\eff_w} \rightsquigarrow \err{\eff_r}{\eff_w}:
  \emph{Err}}[\rulename{T-Err}]

  \inference{\tenv(\var) = \type}{\Sigma, \tenv \vdash_
  {\classtable} \var \rightsquigarrow \var: \type}[
  \rulename{T-Var}]
  
  \inference{\Sigma, \tenv \vdash_{\classtable} \branch
  \rightsquigarrow \branch': \emph{Bool}}
  {\Sigma, \tenv \vdash_{\classtable}
  !\branch \rightsquigarrow !\branch': \emph{Bool}}[\rulename{T-NegB}]
  
  \inference{\Sigma, \tenv \vdash_{\classtable} \branch_1
  \rightsquigarrow \branch'_1: \emph{Bool}\\
  \Sigma, \tenv \vdash_{\classtable} \branch_2
  \rightsquigarrow \branch'_2: \emph{Bool}}
  {\Sigma, \tenv \vdash_{\classtable}
  \branch_1 \lor \branch_2 \rightsquigarrow
  \branch'_1 \lor \branch'_2: \emph{Bool}}[\rulename{T-OrB}]

  \inference{}{\Sigma, \tenv \vdash_{\classtable}
  \ehole: \type \rightsquigarrow \ehole: \type}[\rulename{T-Hole}]

  \inference{}{\Sigma, \tenv \vdash_{\classtable}
  \effhole: \eff \rightsquigarrow (\effhole: \eff): \emph{Obj}}[\rulename{T-EffHole}]
  
  \inference{\val: \type_1 \in \Sigma &
  \type_1 \leq \type_2}
  {\Sigma, \tenv \vdash_{\classtable} \ehole: \type_2
  \rightsquigarrow \val: \type_1}[\rulename{S-Const}]
  
  \inference{\tenv(\var) = \type_1 &
  \type_1 \leq \type_2}
  {\Sigma, \tenv \vdash_{\classtable} \ehole: \type_2
  \rightsquigarrow \var: \type_1}[\rulename{S-Var}]

  \inference{\meth: \mthtype{\type_1}{\type_2} \in \classtable(\class) &
  \type_2 \leq \type_3}
  {\Sigma, \tenv \vdash_{\classtable} \ehole:\type_3
  \rightsquigarrow \emethcall{(\ehole: \class)}{\meth}{\ehole: \type_1}: \type_2}[\rulename{S-App}]
  
  \inference{\Sigma, \tenv \vdash_{\classtable}
  \expr_1 \rightsquigarrow \expr_1': \type_1&
  \Sigma, \tenv \vdash_{\classtable} \expr_2
  \rightsquigarrow \expr_2': \type_2}
  {\Sigma, \tenv \vdash_{\classtable} 
  \eseq{\expr_1}{\expr_2} \rightsquigarrow \eseq{\expr_1'}{\expr_2'}:
  \type_2}[\rulename{T-Seq}]
  
  \inference{\Sigma, \tenv \vdash_{\classtable}
  \expr_1 \rightsquigarrow \expr_1': \type&
  \Sigma, \tenv \vdash_{\classtable} \expr_2
  \rightsquigarrow \expr_2': \type_3\\
  \meth: \mthtype{\type_1}{\type_2} \in \classtable(\class)&
  \type \leq \class & \type_3 \leq \type_1}
  {\Sigma, \tenv \vdash_{\classtable}
  \emethcall{\expr_1}{\meth}{\expr_2} \rightsquigarrow
  \emethcall{\expr_1'}{\meth}{\expr_2'}: \type_2}[\rulename{T-App}]
  
  \inference{\Sigma, \tenv \vdash_{\classtable} \expr_1
  \rightsquigarrow \expr'_1: \type_1\\
  \Sigma, \tenv[\var \mapsto \type_1] \vdash_{\classtable} \expr_2
  \rightsquigarrow \expr'_2: \type_2}
  {\Sigma, \tenv \vdash_{\classtable}
  \elet{\var}{\expr_1}{\expr_2} \rightsquigarrow
  \elet{\var}{\expr'_1}{\expr'_2}: \type_2}[\rulename{T-Let}]
  
  \inference{\Sigma, \tenv \vdash_{\classtable} \branch
  \rightsquigarrow \branch': \emph{Bool}\\
  \Sigma, \tenv \vdash_{\classtable} \expr_1
  \rightsquigarrow \expr'_1: \type_1&
  \Sigma, \tenv \vdash_{\classtable} \expr_2
  \rightsquigarrow \expr'_2: \type_2}
  {\Sigma, \tenv \vdash_{\classtable}
  \eif{\branch}{\expr_1}{\expr_2} \rightsquigarrow 
  \eif{\branch'}{\expr'_1}{\expr'_2}: \type_1 \cup \type_2}[\rulename{T-If}]
  
  \end{mathpar}
  \caption{Type checking and type-directed synthesis rules}
  \label{fig:apdx-type-synthesis-rules}
\end{figure*}

Figure \ref{fig:apdx-type-synthesis-rules} shows all the type checking and
type-directed synthesis rules. The repeated rules are same as
\S~\ref{sec:formalism}. \rulename{T-Nil} type checks the value \vnil and
assigns it the type \emph{Nil}. The rules \rulename{T-True}, \rulename{T-False},
\rulename{T-Obj} and \rulename{T-Err} do the same for the values \vtrue,
\vfalse, \classobj{\class} and \err{\eff_r}{\eff_w} respectively.
Similarly \rulename{T-NegB} and \rulename{T-OrB} give the rules to type check
conditionals that contain negation or disjunction. \rulename{T-EffHole}
typechecks an effect hole with the \emph{Obj} type. It can be narrowed
to a more precise type when synthesis rules are applied. \rulename{T-Seq} does
type checking and synthesis for sequences and \rulename{T-App} type
checks or synthesizes terms in the receiver and arguments of the method
call. \rulename{T-If} type checks if-else expressions.

\begin{figure*}
  \centering
  \begin{mathpar}
  \begin{array}{llcll}
  \hashole
  & \eseq{\expr_1}{\expr_2}
  & = & \hashole\ \expr_1 \land \hashole\ \expr_2&\\
  \hashole
  & \emethcall{\expr_1}{\meth}{\expr_2}
  & = & \hashole\ \expr_1 \land \hashole\ \expr_2&\\
  \hashole
  & \ehole: \type
  & = & \vfalse&\\
  \hashole
  & \elet{\var}{\expr_1}{\expr_2}
  & = & \hashole\ \expr_1 \land \hashole\ \expr_2&\\
  \hashole
  & \eif{\branch}{\expr_1}{\expr_2}
  & = & \hashole\ \branch \land \hashole\ \expr_1 \land \hashole\ \expr_2&\\
  \hashole
  & !\branch
  & = & \hashole\ \branch&\\
  \hashole
  & \branch_1 \lor \branch_2
  & = & \hashole\ \branch_1 \land \hashole\ \branch_2&\\
  \hashole
  & \_
  & = & \vtrue&\\
  \end{array}
  \end{mathpar}


  \begin{mathpar}
  \begin{array}{llcll}
  \size
  & \eseq{\expr_1}{\expr_2}
  & = & \size\ \expr_1 + \size\ \expr_2&\\
  \size
  & \emethcall{\expr_1}{\meth}{\expr_2}
  & = & \size\ \expr_1 + \size\ \expr_2 + 1&\\
  \size
  & \elet{\var}{\expr_1}{\expr_2}
  & = & \size\ \expr_1 + \size\ \expr_2&\\
  \size
  & \eif{\branch}{\expr_1}{\expr_2}
  & = & \size\ \branch + \size\ \expr_1 + \size\ \expr_2&\\
  \size
  & !\branch
  & = & \size\ \branch&\\
  \size
  & \branch_1 \lor \branch_2
  & = & \size\ \branch_1 + \size\ \branch_2&\\
  \size
  & \_
  & = & 0&\\
  \end{array}
  \end{mathpar}
  \caption{Helper functions used by \name}
  \label{fig:aux-func}
\end{figure*}

\subsection{Algorithm}
\label{subsec:appendix-algorithm}

\begin{algorithm}
\caption{Synthesis of programs that passes a spec $s$}
\label{alg:syn-loop}
\begin{algorithmic}[1]
\Procedure{Generate}{\mthtype{\type_1}{\type_2}, \classtable, $\Sigma$, s, maxSize}

  \State \tenv $\gets \lbrack \var \mapsto \type_1\rbrack$

  \State $\expr_0 \gets \ehole: \type_2$

  \State workList $\gets \lbrack (0, \expr_0)
  \rbrack$

  \While{workList is not empty}

    \State $(c, \expr_b) \gets \textrm{pop(workList)}$

    \State $\omega_\type$ $\gets \{ (c, \expr_t) \mid
    \Sigma, \tenv \vdash \expr_b \rightsquigarrow \expr_t: \type \}$

    \State $\omega_{eval} \gets \{ (c, \expr_t) \in
    \omega_\type \land \hashole(\expr_t)\}$

    \State $\omega_r \gets \omega_\type - \omega_{eval}$

    \ForAll{$(c, \expr_t) \in \omega_{eval}$}

      \State $c_r, \val_r \gets \textsc{EvalProgram}(\expr_t, s)$

      \If{$\val_r = \err{\eff_r}{\eff_w}$}

        \State $\omega_r \gets \omega_r \cup \{ (c, \expr_f) \mid
        \Sigma, \tenv, \eff_r \vdash \expr_t \twoheadrightarrow \expr_f \}$
      \Else

        \State \Return{$\expr_t$}

      \EndIf
    \EndFor


    \State $\omega_r \gets \{ (c, \expr_b) \in
    \omega_r \land \size(\expr_b) \leq \textrm{maxSize}\}$

    \State workList $\gets$ reorder(workList + $\omega_r$)

  \EndWhile

  \State \Return Error: No solution found
\EndProcedure
\\
\Procedure{EvalProgram}{\expr, $\langle \precond, \postcond
\rangle$}
  \State $\program \gets \eprog{\meth}{\var}{\expr}$, $E \gets \lbrack \rbrack$
  \State $(c, \mathcal{R}) \gets \llbracket E, 0,
  \rwpair{\pure}{\pure},
  \eseq{\program}{\eseq{\precond}{\postcond}} \rrbracket \hookrightarrow^*_{\classtable} \llbracket E',
  c, \rwpair{\eff_r}{\eff_w}, \mathcal{R} \rrbracket$
  \State \Return $(c, \mathcal{R})$
\EndProcedure
\end{algorithmic}
\end{algorithm}

Algorithm~\ref{alg:syn-loop} describes the synthesis of candidates that pass a
single spec. The algorithm uses a work
list, which initially contains a tuple with the number of passed assertions starting at 0
initially and the initial hole of type $\type_2$. The first
tuple is popped off the work list and applies type or effect guided
synthesis rules, the $\rightsquigarrow$ relation, to a base expression $\expr_b$ to
build the set of new expressions $\omega_\type$. Next, expressions
with no holes are filtered into a list of 
\emph{evaluable} expressions $\omega_{eval}$. Then, \textsc{EvalProgram}
is called to make a program \eprog{\meth}{\var}
{\expr_t} and evaluate spec $s$ in an environment $E$,
starting from a passed assertion count of 0.

If the evaluation of the postcondition results in an error $\err{\eff_r}
{\eff_w}$, the algorithm proceeds to introduce an effect hole, using the
relation $\twoheadrightarrow$ to build the set $\omega_\eff$. If a program passes
all the assertions then it means a correct solution has been found so,
\textsc{Generate} returns it. Finally, the algorithm collects all the 
\emph{remainder} expressions $\omega_r$ with holes and filters the programs that
exceed the maximum permissible size \emph{maxSize}. This bounds
the search to particular search space size. It then takes the filtered
programs and programs from the remainder of the work list and reorders
them. The programs are sorted by the number of passed assertions $c$ in the
decreasing order and then by the program size in the increasing order. This
assumes that a program that is more likely correct and smaller will be selected
earlier for processing in the work list. Lastly, if \textsc{Generate}
doesn't find a correct program in that search space it will return an
error for the same.

Figure~\ref{fig:aux-func} shows the formal definitions of
\code{hasHole}, and \code{size} functions.

\subsection{Branch pruning rules}
\label{subsec:branch-pruning}

\begin{figure}
\begin{align}
\begin{split}
\langle \expr_1, \branch_1, \specs_1 \rangle \oplus \langle \expr_2, \branch_2, \specs_2 \rangle = \langle \branch_1, \branch_1 \lor \branch_2, \specs_1 \cup \specs_2 \rangle\\
\textnormal{if}\ \expr_1 \equiv \vtrue, \expr_2 \equiv \vfalse\ \textnormal{and}\ \branch_1 \equiv\ !\branch_2\label{eq:merge-rule5}
\end{split}
\\
\begin{split}
\langle \expr_1, \branch_1, \specs_1 \rangle \oplus \langle \expr_2, \branch_2, \specs_2 \rangle = \langle \branch_2, \branch_1 \lor \branch_2, \specs_1 \cup \specs_2 \rangle\\
\textnormal{if}\ \expr_1 \equiv \vfalse, \expr_2 \equiv \vtrue\ \textnormal{and}\ \branch_1 \equiv\ !\branch_2\label{eq:merge-rule6}
\end{split}
\\
\begin{split}
\langle \expr_1, \branch_1, \specs_1 \rangle \oplus \langle \expr_2, \branch_2, \specs_2 \rangle = \langle \expr_1, \branch_1, \specs_1 \rangle \oplus \langle \expr_2, \branch_g, \specs_2 \rangle\ \textnormal{if}\ \branch_g \equiv\ !\branch_1\\
\textnormal{and}\ \forall \langle \precond_i, \postcond_i \rangle \in \specs_2. \eprog{\meth}{\var}{\branch_g} \vdash \precond_i; \eassert{\var_r} \Downarrow \val\label{eq:merge-rule7}
\end{split}
\\
\begin{split}
\langle \expr_1, \branch_1, \specs_1 \rangle \oplus \langle \expr_2, \branch_2, \specs_2 \rangle = \langle \expr_1, \branch_g, \specs_1 \rangle \oplus \langle \expr_2, \branch_2, \specs_2 \rangle\ \textnormal{if}\ \branch_g \equiv\ !\branch_2 \\
\textnormal{and}\ \forall \langle \precond_i, \postcond_i \rangle \in \specs_1. \eprog{\meth}{\var}{\branch_g} \vdash \precond_i; \eassert{!\var_r} \Downarrow \val\label{eq:merge-rule8}
\end{split}
\end{align}
\caption{Branch pruning rules.}
\label{fig:branch-reduction}
\end{figure}

Figure~\ref{fig:branch-reduction} formally describes the rules that
allows \name to do term rewriting in \corelang for branch pruning. These
are useful particularly for reducing boolean programs.
Rules \ref{eq:merge-rule5} and 
\ref{eq:merge-rule6} allows us to rewrite expressions into their
branch condition if the expression body is \vtrue or \vfalse reducing
expressions like \eif{\branch}{\vtrue}{\vfalse} to \branch. Rules 
\ref{eq:merge-rule7} and \ref{eq:merge-rule8} guess a conditional that
is the negation of the other, if the negation holds for the tests.
Any $\oplus$ term where two branch conditions are negations of each
other reflect a shorter program \eif{\branch_1}{\expr_1}{\expr_2} in
\corelang. If it is a boolean program, then it might even enable
application of rules \ref{eq:merge-rule5} and \ref{eq:merge-rule6}
producing a single line program.

\end{document}